\documentclass[aps,prd,showpacs,reprint]{revtex4-1}
\usepackage{graphicx}
\usepackage{dcolumn}
\usepackage{amsmath}
\usepackage{xcolor}
\usepackage{epsfig}
\RequirePackage{xspace}
\usepackage{relsize}

\begin{document}
\title{Branching ratio and \textit{CP} violation of $B\rightarrow K\pi$ decays
  in a modified perturbative QCD approach}
\author{Ru-Xuan Wang} \email{wangrx@mail.nankai.edu.cn}
\author{Mao-Zhi Yang} \email{yangmz@nankai.edu.cn}
\affiliation{School of Physics, Nankai University, Tianjin 300071, P.R. China}
\date{\today}
\pacs{12.38.Bx, 12.39.St, 13.25.Hw}

\begin{abstract}
  We calculate the branching ratios and \textit{CP} violations for $B\rightarrow K\pi$ decays in a
  modified perturbative QCD approach based on $k_T$ factorization. The resummation effect of the transverse momentum regulates the endpoint singularity. Using the $B$ meson wave function that is obtained in the relativistic potential model, soft contribution can not be suppressed effectively by Sudakov factor.
  Soft scale cutoff and soft $BK$, $B\pi$ and $K\pi$ form factors have to be introduced. The most important next-to-leading-order contributions from the vertex corrections, the quark loops, and the magnetic penguins are also considered. In addition, the contribution of the color-octet hadronic matrix element is included which is essentially of long-distance dynamics. Our predictions for all the branching ratios and most \textit{CP} violations are well consistent with the experimental data. Especially the theoretical result of dramatic difference between the \textit{CP} violations of $B^+\rightarrow K^+\pi^0$ and $B^0\rightarrow K^+\pi^-$ is in good agreement with experimental measurement, therefore the $K\pi$ puzzle in $B$ decays can be resolved in our way of the modified perturbative QCD approach.
\end{abstract}
\maketitle

\section{Introduction} \label{sec:intro}
Over the past twenty years, there has been a great deal of interest in $B\rightarrow K\pi$ decays.
From $B$ factory experiments, a large amount of data about $B$ decays has been collected.
The precise data revealed significant difference between experiment measurements and theoretical predictions.
For $B\rightarrow K\pi$ decays, the expected \textit{CP} violations of $B^+\rightarrow K^+\pi^0$ and
$B^0\rightarrow K^+\pi^-$ decays are roughly equal from the theoretical point of view \cite{beneke2003qcd,li2005resolution}.
The branching ratios and \textit{CP} violations of $B\rightarrow K\pi$ decays measured by experiment are
\cite{Zyla:2020zbs}
\begin{equation}
  \begin{split}
    &B(B^+\rightarrow K^0\pi^+)=(2.37\pm 0.08)\times 10^{-5},\\
    &B(B^+\rightarrow K^+\pi^0)=(1.29\pm 0.05)\times 10^{-5},\\
    &B(B^0\rightarrow K^+\pi^-)=(1.96\pm 0.05)\times 10^{-5},\\
    &B(B^0\rightarrow K^0\pi^0)=( 9.9\pm  0.5)\times 10^{-6},\\
  \end{split}
\end{equation}
and
\begin{equation}\label{exp-cp}
  \begin{split}
    &A_{CP}(B^+\rightarrow K^0\pi^+)=-0.017\pm 0.016, \\
    &A_{CP}(B^+\rightarrow K^+\pi^0)= 0.037\pm 0.021, \\
    &A_{CP}(B^0\rightarrow K^+\pi^-)=-0.083\pm 0.004,\\
    &A_{CP}(B^0\rightarrow K^0\pi^0)= 0.00 \pm 0.13.  \\
  \end{split}
\end{equation}
One can obtain
$\Delta A_{CP}\equiv A_{CP}(B^+\rightarrow K^+\pi^0)-A_{CP}(B^0\rightarrow K^+\pi^-)=0.120\pm 0.021$ from the data given in Eq. (\ref{exp-cp}).
The difference deviates from zero by more than $5\sigma$. The understanding of the experimental data of branching ratios of $B\to K\pi$ decays is also puzzling \cite{BFRS2003}. This is what is called $B\rightarrow K\pi$ puzzle.

$B\to K\pi$ puzzle attracted a lot of interest from the theoretical point of view. Theoretical analysis shows that  $B\to K\pi$ puzzle may indicate significant enhancement of electroweak (EW) penguin and color-suppressed tree contributions \cite{BFRS2003,KSC2005,FRS2007}.

The original predictions for branching ratios and $CP$ violations of $B\to K\pi$, $\pi\pi$ decays in perturbative QCD (PQCD) approach can be found in Refs. \cite{PQCD1,PQCD2,PQCD3}. To solve the $K\pi$ puzzle in $B$ decays, the next-to-leading-order QCD corrections have been taken into account in PQCD approach in Refs. \cite{li2005resolution,bai2014revisiting}. A soft factor that enhances the nonfactorizable amplitudes has been introduced based on the analysis of soft divergences that appear in higher order loop corrections in QCD and used to solve the $K\pi$ puzzle in Refs.~\cite{li2011possible,li2014,LLX2016}. Ref. \cite{CSYL2014} made effort to understand the puzzling problem in QCD factorization (QCDF) approach \cite{beneke1999qcd,beneke2000qcd,beneke2001qcd} by considering scattering and annihilation contributions. There are also works where new physics effects are considered to solve the $K\pi$ puzzle \cite{BCLL2004,BHLDS2005,ADHO,kim2008analytic,BDLRR2018,datta2019unified}. All of these works can reduce the discrepancy between theoretical prediction and experimental data, and shows positive signals to understand the puzzle, but there is still possibility to further study the $K\pi$ puzzle with a new point of view.

In this work we study $B\rightarrow K\pi$ decays in a modified perturbative QCD (PQCD) approach, with which $B\to\pi\pi$ decays have been studied very recently in Ref. \cite{luyang2022}, where both the branching ratios and $CP$ violation in three $B\to\pi\pi$ decay modes are well consistent with experimental data. It is found that, using $B$ meson wave function that obtained by solving the bound state equation in relativistic potential model \cite{yang2012wave,liu2014spectrum,liu2015spectrum,sun2017decay,sun2019wave}, the suppression to soft contribution from Sudakov factor is not large enough. A soft truncation have to be introduced in an appropriate momentum scale $\mu_c$. When the momentum transfer larger than this critical momentum scale $\mu_c$, the contributions to $B\rightarrow \pi$ and $B\rightarrow K$ transition form factors can be calculated perturbatively. And soft form factors need to be introduced to include soft contributions with the momentum transfer lower than the momentum scale of cutoff. With the soft cutoff and soft form factors, the calculation of $B\rightarrow \pi$ and $B\rightarrow K$ transition form factors becomes more reliable.

The branching ratios and \textit{CP} violations of $B\rightarrow K\pi$ decays are calculated in this work. The amplitudes are treated perturbatively when the momentum transfer larger than the soft cutoff scale. We also consider the important next-to-leading-order contributions of hard part from the vertex corrections, the quark loops and the magnetic penguin. As for the soft part with momentum transfer lower than the cutoff scale, we introduce the $BK$ and $B\pi$ transition, and $K\pi$ production soft form factors. These factors are nonperturbative input parameters. To improve the consistency between theoretical calculation and experimental data, we find that the nonzero color-octet matrix element $\langle K\pi|(\bar{s}T^aq)(\bar{q}T^ab)|B\rangle$, which is derived from the analysis of color structure of quark-antiquark current operators, is necessary. With the appropriate input parameters, our prediction of branching ratio and \textit{CP} violation is well consistent with experimental data.

The paper is orginized as follows.
The perturbative calculation of leading order contributions of $B\rightarrow K\pi$ decays are presented
in Sec.~\ref{sec:perturlo}.
The important next-to-leading order contributions are considered in Sec.~\ref{sec:nlo}.
The contributions of nonperturbative parameters are investigated in Sec.~\ref{sec:softff} and \ref{sec:co}.
The numerical results are shown in Sec.~\ref{sec:result}.
We conclude the analysis in Sec.~\ref{sec:sum}.

\section{The hard amplitudes of leading order contributions in perturbative QCD} \label{sec:perturlo}
When the momentum transfer in the transition process is larger than the cutoff scale $\mu_c$,
which is used to seperate the contributions of hard and soft parts,
the decay amplitude can be treated perturbatively.
Typically, the critical scale $\mu_c$ can be approximately taken to be $1.0~\textup{GeV}$.
In perturbative QCD approach, if $B$ meson decays into two light mesons,
the process is dominated by one hard gluon exchanged diagrams.
The decay amplitudes can be arranged as the convolution of hard scattering process and meson wave functions
\begin{equation}
  \begin{split}
    \mathcal{M}=\int d^3k_1 \int d^3k_2 \int d^3k_3 \Phi^B(k_1,\mu) C(\mu) \\
    \quad \times H(k_1,k_2,k_3,\mu) \Phi^\pi(k_2,\mu) \Phi^K(k_3,\mu),\\
  \end{split}
\end{equation}
where $H$ contains the hard scattering dynamics which is calculable using perturbation theory,
$C(\mu)$'s are Wilson coefficients, $\Phi(x)^{B,\pi,K}$ are meson light-cone distribution amplitudes that absorb nonperturbative interactions related to meson states.

For the $b\rightarrow s$ transtion, the effective Hamiltonian is given by \cite{buchalla1996weak}
\begin{equation}
  \begin{split}
    H_{\textup{eff}}=&\frac{G_F}{\sqrt{2}}\biggl[V_u(C_1O_1^u+C_2O_2^u) -V_t\biggl(\sum_{i=3}^{10}C_iO_i \\
      &+C_{8g}O_{8g}\biggr)\biggr], \\
  \end{split}
\end{equation}
where $V_u=V_{ub}V_{us}^*$ and $V_t=V_{tb}V_{ts}^*$, are Cabibbo-Kobayashi-Maskawa (CKM) matrix elements, $G_F=1.16639\times 10^{-5}~\textup{GeV}^{-2}$ Fermi constant, and the $C_i$'s Wilson coefficients. The operators $O_i$ in the effective Hamiltonian are
\begin{equation}\label{eq:o1}
  \begin{split}
    &O_1^u=(\bar{s}_\alpha\gamma^\mu(1-\gamma_5)u_\beta)(\bar{u}_{\beta}\gamma_\mu(1-\gamma_5)b_\alpha),\\
    &O_2^u=(\bar{s}_\alpha\gamma^\mu(1-\gamma_5)u_\alpha)(\bar{u}_{\beta}\gamma_\mu(1-\gamma_5)b_\beta),\\
  \end{split}
\end{equation}
\begin{equation}\label{eq:o3}
  \begin{split}
    &O_3=(\bar{s}_\alpha\gamma^\mu(1-\gamma_5)b_\alpha)
      \sum_{q^\prime}(\bar{q}_{\beta}^\prime\gamma_\mu(1-\gamma_5)q_\beta^\prime),\\
    &O_4=(\bar{s}_\alpha\gamma^\mu(1-\gamma_5)b_\beta)
      \sum_{q^\prime}(\bar{q}_{\beta}^\prime\gamma_\mu(1-\gamma_5)q_\alpha^\prime),\\
    &O_5=(\bar{s}_\alpha\gamma^\mu(1-\gamma_5)b_\alpha)
      \sum_{q^\prime}(\bar{q}_{\beta}^\prime\gamma_\mu(1+\gamma_5)q_\beta^\prime),\\
    &O_6=(\bar{s}_\alpha\gamma^\mu(1-\gamma_5)b_\beta)
      \sum_{q^\prime}(\bar{q}_{\beta}^\prime\gamma_\mu(1+\gamma_5)q_\alpha^\prime),\\
  \end{split}
\end{equation}
\begin{equation}\label{eq:o7}
  \begin{split}
    &O_7=\frac{3}{2}(\bar{s}_\alpha\gamma^\mu(1-\gamma_5)b_\alpha)
      \sum_{q^\prime}e_{q^\prime}(\bar{q}_{\beta}^\prime\gamma_\mu(1+\gamma_5)q_\beta^\prime),\\
    &O_8=\frac{3}{2}(\bar{s}_\alpha\gamma^\mu(1-\gamma_5)b_\beta)
      \sum_{q^\prime}e_{q^\prime}(\bar{q}_{\beta}^\prime\gamma_\mu(1+\gamma_5)q_\alpha^\prime),\\
    &O_9=\frac{3}{2}(\bar{s}_\alpha\gamma^\mu(1-\gamma_5)b_\alpha)
      \sum_{q^\prime}e_{q^\prime}(\bar{q}_{\beta}^\prime\gamma_\mu(1-\gamma_5)q_\beta^\prime),\\
    &O_{10}=\frac{3}{2}(\bar{s}_\alpha\gamma^\mu(1-\gamma_5)b_\beta)
      \sum_{q^\prime}e_{q^\prime}(\bar{q}_{\beta}^\prime\gamma_\mu(1-\gamma_5)q_\alpha^\prime),\\
  \end{split}
\end{equation}
\begin{equation}\label{eq:o8g}
  \begin{split}
    O_{8g}=\frac{g_s}{8\pi^2}m_b\bar{s}_\alpha\sigma^{\mu\nu}(1+\gamma_5)
      T^a_{\alpha\beta}G^a_{\mu\nu}b_\beta,
  \end{split}
\end{equation}
where $\alpha$ and $\beta$ are the color indices. The summation of $q^\prime$ runs through $u$, $d$, $s$, $c$, and $b$ quarks.

The matrix element $\langle 0|\bar{q}_\beta(z)[z,0]b_\alpha(0)|\bar{B}\rangle$ can be used to define the
$B$ meson wave function
\begin{equation}
  \langle 0|\bar{q}_\beta(z)[z,0]b_\alpha(0)|\bar{B}\rangle
    =\int d^3{k}\Phi_{\alpha\beta}^B(\vec{k})\exp(-ik\cdot z),
\end{equation}
where $[z,0]$ represents the path-ordered exponential
\begin{equation}
  [z,0]=\mathcal{P}\exp\left[-ig_sT^a\int_0^1d\alpha z^\mu A_\mu^a(\alpha z)\right].
\end{equation}
We use the $B$ wave function that is obtained by solving bound-state equation in the QCD-inspired
relativistic potential model \cite{yang2012wave,liu2014spectrum,liu2015spectrum,sun2017decay}, where the mass spectrum and decay constants of $b$-flavored meson system calculated simultaneously with the wave functions
are consistent with experimental data.

In the rest-frame of $B$ meson, the spinor wave function $\Phi_{\alpha\beta}^B(\vec{k})$ is given by
\cite{sun2017decay}
\begin{equation}
  \begin{split}
    \Phi_{\alpha\beta}^B(\vec{k})&=\frac{-if_Bm_B}{4}K(\vec{k}) \\
    &\times\biggl\{(E_Q+m_Q)\frac{1+\not{v}}{2}
      \biggl[\left(\frac{k_+}{\sqrt{2}}+\frac{m_q}{2}\right)\not{n}_+ \\
    &\quad+\left(\frac{k_-}{\sqrt{2}}+\frac{m_q}{2}\right)\not{n}_-
      -k_\perp^\mu\gamma_\mu\biggr]\gamma_5 \\
    &\quad-(E_q+m_q)\frac{1-\not{v}}{2}
      \biggl[\left(\frac{k_+}{\sqrt{2}}-\frac{m_q}{2}\right)\not{n}_+ \\
    &\quad+\left(\frac{k_-}{\sqrt{2}}-\frac{m_q}{2}\right)\not{n}_-
      -k_\perp^\mu\gamma_\mu\biggr]\gamma_5\biggr\}_{\alpha\beta},\\
  \end{split}
\end{equation}
where $Q$ and $q$ represent the heavy and light quarks in $B$ meson $(b\bar{q})$, respectively. $v$ is the four-speed of $B$ meson which satisfies $p_B^\mu = m_B v^\mu$ and $v^\mu=(1,0,0,0)$, and $k$ is the momentum of the light quark in the rest frame of the meson. $k^\pm$ and $k_\perp$ are defined by
\begin{equation}
  k^\pm=\frac{E_q\pm k^3}{\sqrt{2}},\quad k_\perp^\mu=(0,k^1,k^2,0).
\end{equation}
$n^\mu_\pm$ are two light-like vectors with $n^\mu_\pm = (1,0,0,\mp1)$,
and $K(\vec{k})$ is the function proportional to the $B$-meson wave function
\begin{equation}
  K(\vec{k}) = \frac{2N_B\Psi_0(\vec{k})}{\sqrt{E_qE_Q(E_q+m_q)(E_Q+m_Q)}},
\end{equation}
with the normalization constant $N_B=\frac{1}{f_B}\sqrt{\frac{3}{(2\pi)^3m_B}}$ and
the $B$ meson wave function
\begin{equation}
  \Psi_0(\vec{k}) = a_1 e^{a_2 |\vec{k}|^2 + a_3 |\vec{k}| +a_4},
\end{equation}
where the parameters are \cite{sun2017decay}
\begin{equation}
  \begin{split}
    &a_1=4.55_{-0.30}^{+0.40}~\textup{GeV}^{-3/2},\\
    &a_2=-0.39_{-0.20}^{+0.15}~\textup{GeV}^{-2},\\
    &a_3=-1.55\pm 0.20~\textup{GeV}^{-1},\\
    &a_4=-1.10_{-0.05}^{+0.10}.\\
  \end{split}
\end{equation}

\begin{figure}[b]
  \includegraphics[width=0.4\textwidth]{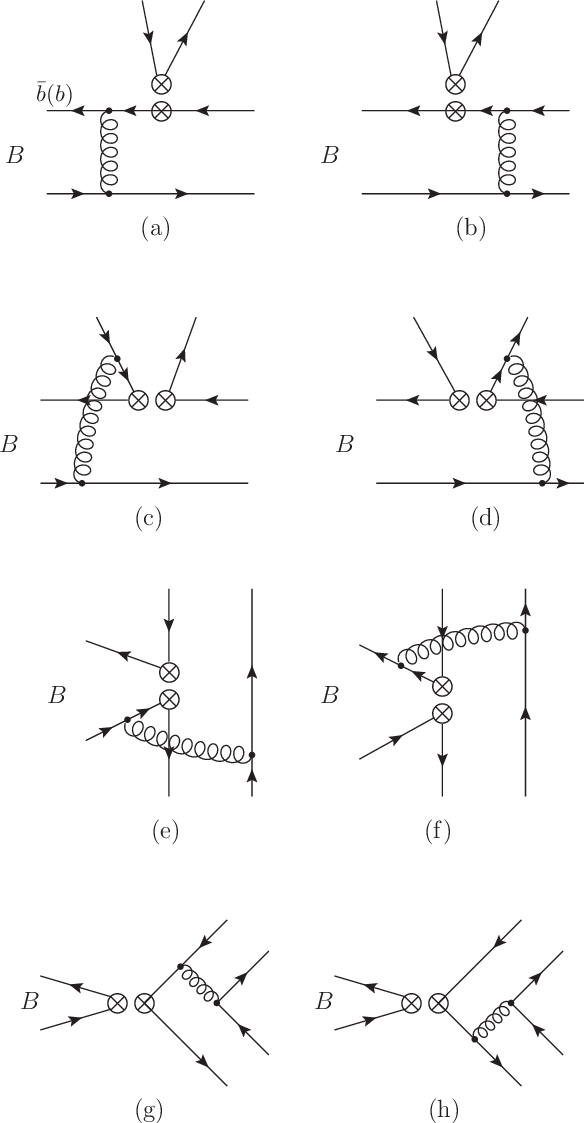}
  \caption{\label{fig:diagrams} Diagrams which contribute to the $B\rightarrow K\pi$ decays.}
\end{figure}
In $B\rightarrow K\pi$ decays, the mass difference of $B$ and final state mesons is large.
The momenta of the outgoing $K$ and $\pi$ mesons are large, so the wave functions of light mesons can be defined on the light-cone \cite{braun1990conformal,ball1999theoretical,ball2006higher}
\begin{equation} \label{pion-wave}
  \begin{split}
    &\langle\pi(p)|\bar{q}_\delta(x)q_\gamma^\prime(0)|0\rangle \\
    &\quad=\int dud^2k_{q\perp}\Phi_{\gamma\delta}^\pi
      \exp\left[i(up\cdot x-x_\perp\cdot k_{q\perp})\right], \\
  \end{split}
\end{equation}
with the spinor wave function $\Phi_{\gamma\delta}^{\pi}$ being
\begin{equation} \label{pion-spinor-wave}
  \begin{split}
    \Phi_{\gamma\delta}^{\pi}&=\frac{if_\pi}{4}
      \biggl[\not{p}\gamma_5\phi_\pi(u,k_{q\perp})-\mu_\pi\gamma_5\phi_P^\pi(u,k_{q\perp}) \\
    &\quad +\mu_\pi\gamma_5\sigma^{\mu\nu}p_\mu z_\nu\frac{\phi_\sigma^\pi(u,k_{q\perp})}{6}
      \biggr]_{\gamma\delta}, \\
  \end{split}
\end{equation}
where $f_\pi$ is pion decay constant, and $\mu_\pi$ is the chiral parameter with
\begin{equation}
  \mu_\pi=m_\pi^2/(m_u+m_d).
\end{equation}
$\phi_\pi$, $\phi_P^\pi$ and $\phi_\sigma^\pi$ are the twist-2 and twist-3 light-cone distribution amplitudes.
In momentum space, the pion wave function can be expressed as
\begin{equation}
  \begin{split}
    \Phi_{\gamma\delta}^{\pi}&=\frac{if_\pi}{4}
      \biggl[\not{p}\gamma_5\phi_\pi(u,k_{q\perp})-\mu_\pi\gamma_5\phi_P^\pi(u,k_{q\perp}) \\
    &\quad +\mu_\pi\gamma_5i\sigma^{\mu\nu}\frac{p_\mu\bar{p}_\nu}{p\cdot\bar{p}}
      \frac{\phi_\sigma^{\prime\pi}(u,k_{q\perp})}{6} \\
    &\quad -\mu_\pi\gamma_5i\sigma^{\mu\nu}p_\mu\frac{\phi_\sigma^\pi(u,k_{q\perp})}{6}
      \frac{\partial}{\partial k_{q\perp\nu}}\biggr]_{\gamma\delta}, \\
  \end{split}
\end{equation}
where $\phi_\sigma^{\prime\pi}(u,k_{q\perp})=\frac{\partial\phi_\sigma^\pi(u,k_{q\perp})}{\partial u}$, $\bar{p}$ the momentum with the moving direction opposite to that of pion and the energy the same.
For the wave function of $K$ meson, one can get it by just replacing the distribution amplitudes of pion with that
of the kaon \cite{ball2006higher}.

There are eight diagrams contributing to $B\rightarrow K\pi$ decays in leading order (LO) in QCD which are shown in Fig.~\ref{fig:diagrams}. During the calculation, we keep the transverse momentum of quarks and gluons. At the endpoint region, i.e., when the momentum fraction of parton $x\rightarrow 0$, the transverse momentum cannot be neglected. We encounter double logarithm divergence such as $\alpha_s(\mu)\ln^2(k_\perp/\mu)$ when soft and collinear divergences overlap. This large double logarithms should be resummed into the Sudakov factor \cite{li1996perturbative,li1996pqcd}. In addition, there are other double logarithms such as $\alpha_s(\mu)\ln^2x$ from the QCD corrections of the weak vertex. This double logarithm can also be resummed into the threshold factor \cite{li2002threshold}. The Sudakov factor and threshold factor suppress the endpoint singularity and improve the reliability of the calculation of $B$ decays in PQCD approach. For convenience, we perform the calculation of the decay amplitudes in $b$-space where $b$ is the conjugate variable of transverse momentum $k_\perp$.

The diagrams (a) and (b) in Fig.~\ref{fig:diagrams} are factorizable diagrams. Fig.~\ref{fig:diagrams} (c) and (d) are nonfactorizable diagrams, (e) and (f) nonfactorizable annihilation diagrams, and (g) and (h) factorizable annihilation diagrams. The symbol of the circled times in these diagrams stands for four-quark operator insertion which are given in Eqs.~\eqref{eq:o1}--\eqref{eq:o8g}.


Firstly, we calculate the diagrams (a) and (b). If the meson which is factorized out is kaon in diagrams (a) and (b), the contribution with the $(V-A)(V-A)$ operators inserted is
\begin{widetext}
\begin{equation} \label{eq:fe}
  \begin{split}
    F_{e}=&2\pi^2f_Bf_\pi m_B^2\frac{C_F}{N_c}
      \int k_{1\perp}dk_{1\perp}\int_{x_1^d}^{x_1^u}dx_1\int_0^1dx_3
      \int_0^\infty b_1db_1b_3db_3
      \left(\frac{1}{2}m_B+\frac{|\mathbf{k}_{1\perp}|^2}{2x_1^2m_B}\right)
      K(\vec{k}_1)(E_Q+m_Q) \\
    &\times J_0(k_{1\perp}b_1)
      \biggl\{\alpha_s(t_e^1)\biggl[\Bigl((x_3-2)E_q-x_3k_1^3\Bigr)\phi_\pi(x_3,b_3)
      +r_\pi\Bigl((1-2x_3)E_q+k_1^3\Bigr)\phi_{P}^\pi(x_3,b_3)
      -\frac{1}{6}r_\pi\\ &\times\Bigl((1-2x_3)E_q+k_1^3\Bigr)\phi_\sigma^{\prime\pi}(x_3,b_3)\biggr]
      h_e(x_1,1-x_3,b_1,b_3)S_t(x_3)\exp[-S_B(t_e^1)-S_\pi(t_e^1)]
      +\alpha_s(t_e^2)2r_\pi \\ &\times(-E_q+k_1^3)\phi_P^\pi(x_3,b_3)
      h_e(1-x_3,x_1,b_3,b_1)S_t(x_1)\exp[-S_B(t_e^2)-S_\pi(t_e^2)]\biggr\}. 
  \end{split}
\end{equation}
The contribution related to the $(S+P)(S-P)$ operators which come from Fierz transformation of $(V-A)(V+A)$ operators is
\begin{equation} \label{eq:fep}
  \begin{split}
    F_{e}^P=&2\pi^2 f_Bf_\pi m_B^2\frac{C_F}{N_c}\int k_{1\perp}dk_{1\perp}
      \int_{x_1^d}^{x_1^u}dx_1\int_0^1dx_3
      \int_0^\infty b_1db_1b_3db_3
      \left(\frac{1}{2}m_B+\frac{|\mathbf{k}_{1\perp}|^2}{2x_1^2m_B}\right)
      K(\vec{k}_1)(E_Q+m_Q) \\
    &\times J_0(k_{1\perp}b_1)2r_K
      \biggl\{\alpha_s(t_e^1)\biggl[-(E_q+k_1^3)\phi_\pi(x_3,b_3)
      +r_\pi\Bigl((x_3-3)E_q+(1-x_3)k_1^3\Bigr)\phi_P^\pi(x_3,b_3)
      +\frac{1}{6}r_\pi \\ &\times\Bigl((x_3-1)E_q-(1+x_3)k_1^3\Bigr)\phi_\sigma^{\prime\pi}(x_3,b_3)\biggr]
      h_e(x_1,1-x_3,b_1,b_3)S_t(x_3)\exp[-S_B(t_e^1)-S_\pi(t_e^1)]\\
    &+\alpha_s(t_e^2)2r_\pi(-E_q+k_1^3)\phi_P^\pi(x_3,b_3)
      h_e(1-x_3,x_1,b_3,b_1)S_t(x_1)\exp[-S_B(t_e^2)-S_\pi(t_e^2)]\biggr\}.
  \end{split}
\end{equation}
If the meson factorized out is pion in diagrams (a) and (b),
the contributions from this two diagrams with the $(V-A)(V-A)$ and $(V-A)(V+A)$ operators inserted are
\begin{equation} \label{eq:fek}
  \begin{split}
    F_{eK}=&2\pi^2f_Bf_Km_B^2\frac{C_F}{N_c}\int k_{1\perp}dk_{1\perp}
      \int_{x_1^d}^{x_1^u}dx_1\int_0^1dx_2
      \int_0^\infty b_1db_1b_2db_2
      \left(\frac{1}{2}m_B+\frac{|\mathbf{k}_{1\perp}|^2}{2x_1^2m_B}\right)
      K(\vec{k}_1)(E_Q+m_Q)\\
    &\times J_0(k_{1\perp}b_1)
      \biggl\{\alpha_s(t_{eK}^1)
        \biggl[\Bigl((x_2-2)E_q+x_2k_1^3\Bigr)\phi_K(x_2,b_2)
      +r_K\Bigl((1-2x_2)E_q-k_1^3\Bigr)\phi_P^K(x_2,b_2)
      -\frac{1}{6}r_K \\ &\times\Bigl((1-2x_2)E_q-k_1^3\Bigr)\phi_\sigma^{\prime K}(x_2,b_2)\biggr]
      h_e(x_1,1-x_2,b_1,b_2)S_t(x_2)\exp[-S_B(t_{eK}^1)-S_K(t_{eK}^1)]
      +\alpha_s(t_{eK}^2) \\ &\times 2r_K(-E_q-k_1^3)\phi_P^K(x_2,b_2)
      h_e(1-x_2,x_1,b_2,b_1)S_t(x_1)\exp[-S_B(t_{eK}^2)-S_K(t_{eK}^2)]\biggr\},
  \end{split}
\end{equation}
and
\begin{equation} \label{eq:fekp}
  F_{eK}^P=-F_{eK}.
\end{equation}
There are also factorizable annihilation diagrams (g) and (h), where the $B$ meson is factored out.
The results of diagrams (g) and (h) are
\begin{equation} \label{eq:fa}
  \begin{split}
    F_{a}=&2\pi f_Kf_\pi m_B^2\frac{C_F}{N_c}
      \int_0^1dx_2dx_3\int_0^\infty b_2db_2b_3db_3
      \biggl\{\alpha_s(t_a^1)
        \biggl[x_3\phi_K(x_2,b_2)\phi_\pi(x_3,b_3)
        +2r_Kr_\pi(1+x_3)\phi_P^K(x_2,b_2) \\ & \times\phi_P^\pi(x_3,b_3)
    -\frac{1}{3}r_Kr_\pi(1-x_3)\phi_P^K(x_2,b_2)\phi_\sigma^{\prime\pi}(x_3,b_3)\biggr]
      h_a(1-x_2,x_3,b_2,b_3)S_t(x_3)\exp[-S_K(t_a^1)-S_\pi(t_a^1)]\\
    &+\alpha_s(t_a^2)\biggl[-(1-x_2)\phi_K(x_2,b_2)\phi_\pi(x_3,b_3)
      -2r_Kr_\pi(2-x_2)\phi_P^K(x_2,b_2)\phi_P^\pi(x_3,b_3)
      -\frac{1}{3}r_Kr_\pi x_2\phi_\sigma^{\prime K}(x_2,b_2)\\ &\times\phi_P^\pi(x_3,b_3)\biggr]
      h_a(x_3,1-x_2,b_3,b_2)S_t(x_2)\exp[-S_K(t_a^2)-S_\pi(t_a^2)]\biggr\},
  \end{split}
\end{equation}
\begin{equation} \label{eq:fap}
  \begin{split}
    F_{a}^P=&4\pi f_Kf_\pi m_B^2\frac{C_F}{N_c}\chi_B
      \int_0^1dx_2dx_3\int_0^\infty b_2db_2b_3db_3
      \biggl\{\alpha_s(t_a^1)
        \biggl[-r_\pi x_3\phi_K(x_2,b_2)\phi_P^\pi(x_3,b_3)
        +\frac{1}{6}r_\pi x_3\phi_K(x_2,b_2) \\ & \times\phi_\sigma^{\prime\pi}(x_3,b_3)
    -2r_K\phi_P^K(x_2,b_2)\phi_\pi(x_3,b_3)\biggr]
      h_a(1-x_2,x_3,b_2,b_3)S_t(x_3)\exp[-S_K(t_a^1)-S_\pi(t_a^1)]
      +\alpha_s(t_a^2)\\ &\times\biggl[-2r_\pi\phi_K(x_2,b_2)\phi_P^\pi(x_3,b_3)
      -r_K(1-x_2)\phi_P^K(x_2,b_2)\phi_\pi(x_3,b_3)
      -\frac{1}{6}r_K(1-x_2)\phi_\sigma^{\prime K}(x_2,b_2)\phi_\pi(x_3,b_3)\biggr] \\
    &\times h_a(x_3,1-x_2,b_3,b_2)S_t(x_2)\exp[-S_K(t_a^2)-S_\pi(t_a^2)]\biggr\}.
  \end{split}
\end{equation}

As for the nonfactorizable diagrams (c), (d), (e) and (f),
the amplitudes involve all three meson wave functions.
The integral over $b$ using $\delta$ function is necessary.
The amplitudes of the nonfactorizable emission diagrams (c) and (d) are
\begin{equation} \label{eq:me}
  \begin{split}
    \mathcal{M}_{e}=&2\pi^2f_Bf_Kf_\pi m_B^2\frac{C_F}{N_c}
      \int k_{1\perp}dk_{1\perp}\int_{x_1^d}^{x_1^u}dx_1\int_0^1dx_2dx_3
      \int_0^\infty b_1db_1b_2db_2
      \left(\frac{1}{2}m_B+\frac{|\mathbf{k}_{1\perp}|^2}{2x_1^2m_B}\right)
      K(\vec{k}_1)(E_Q+m_Q)\\
    &\times J_0(k_{1\perp}b_1)\phi_K(x_2,b_2)
      \biggl\{\alpha_s(t_d^1)
        \biggl[-x_2(E_q+k_1^3)\phi_\pi(x_3,b_1)
        +r_\pi(1-x_3)(E_q-k_1^3)\phi_P^\pi(x_3,b_1)
    +\frac{1}{6}r_\pi(1-x_3)\\ &\times(E_q-k_1^3)\phi_\sigma^{\prime\pi}(x_3,b_1)\biggr]
      h_d(x_1,x_2,1-x_3,b_1,b_2)\exp[-S_B(t_d^1)-S_K(t_d^1)-S_\pi(t_d^1)|_{b_3\rightarrow b_1}]
    +\alpha_s(t_d^2)\\ &\times
      \biggl[\Bigl((2-x_2-x_3)E_q-(x_2-x_3)k_1^3\Bigr)\phi_\pi(x_3,b_1)
      -r_\pi(1-x_3)(E_q+k_1^3)\phi_P^\pi(x_3,b_1)
    +\frac{1}{6}r_\pi(1-x_3)(E_q+k_1^3)\\ &\times\phi_\sigma^{\prime\pi}(x_3,b_1)\biggr]
      h_d(x_1,1-x_2,1-x_3,b_1,b_2)\exp[-S_B(t_d^2)-S_K(t_d^2)-S_\pi(t_d^2)|_{b_3\rightarrow b_1}]\biggr\},
  \end{split}
\end{equation}
\begin{equation} \label{eq:mep}
  \begin{split}
    \mathcal{M}_{e}^P=&2\pi^2f_Bf_Kf_\pi m_B^2\frac{C_F}{N_c}
      \int k_{1\perp}dk_{1\perp}\int_{x_1^d}^{x_1^u}dx_1\int_0^1dx_2dx_3
      \int_0^\infty b_1db_1b_2db_2
      \left(\frac{1}{2}m_B+\frac{|\mathbf{k}_{1\perp}|^2}{2x_1^2m_B}\right)
      K(\vec{k}_1)(E_Q+m_Q)\\
    &\times J_0(k_{1\perp}b_1)r_K
      \biggl\{\alpha_s(t_d^1)
        \biggl[-x_2(E_q+k_1^3)\phi_P^K(x_2,b_2)\phi_\pi(x_3,b_1)
          -r_\pi\Bigl((1+x_2-x_3)E_q-(1-x_2-x_3)k_1^3\Bigr) \\ & \times\phi_P^K(x_2,b_2)\phi_P^\pi(x_3,b_1)
    -\frac{1}{6}r_\pi\Bigl(-(1-x_2-x_3)E_q
      +(1+x_2-x_3)k_1^3\Bigr)\phi_P^K(x_2,b_2)\phi_\sigma^{\prime\pi}(x_3,b_1)
    +\frac{1}{6}x_2\\ &\times(E_q+k_1^3)\phi_\sigma^{\prime K}(x_2,b_2)\phi_\pi(x_3,b_1)
      +\frac{1}{6}r_\pi\Bigl(-(1-x_2-x_3)E_q
        +(1+x_2-x_3)k_1^3\Bigr)\phi_\sigma^{\prime K}(x_2,b_2)\phi_P^\pi(x_3,b_1)\\
    &+\frac{1}{36}r_\pi\Bigl((1+x_2-x_3)E_q
      -(1-x_2-x_3)k_1^3\Bigr)\phi_\sigma^{\prime K}(x_2,b_2)\phi_\sigma^{\prime\pi}(x_3,b_1)\biggr]
    h_d(x_1,x_2,1-x_3,b_1,b_2)\\
     &\times\exp[-S_B(t_d^1)-S_K(t_d^1)-S_\pi(t_d^1)|_{b_3\rightarrow b_1}]
    +\alpha_s(t_d^2)
      \biggl[(1-x_2)(E_q+k_1^3)\phi_P^K(x_2,b_2)\phi_\pi(x_3,b_1)
        +r_\pi\\ &\times\Bigl((2-x_2-x_3)E_q
          -(x_2-x_3)k_1^3\Bigr)\phi_P^K(x_2,b_2)\phi_P^\pi(x_3,b_1)
    +\frac{1}{6}r_\pi\Bigl(-(x_2-x_3)E_q
      +(2-x_2-x_3)k_1^3\Bigr) \\
    &\times\phi_P^K(x_2,b_2)\phi_\sigma^{\prime\pi}(x_3,b_1)
      +\frac{1}{6}(1-x_2)(E_q+k_1^3)\phi_\sigma^{\prime K}(x_2,b_2)\phi_\pi(x_3,b_1)
    +\frac{1}{6}r_\pi\Bigl(-(x_2-x_3)E_q 
  \end{split}
 \end{equation}
\begin{equation}
  \begin{split}
   ~~&+(2-x_2-x_3)k_1^3\Bigr)\phi_\sigma^{\prime K}(x_2,b_2)\phi_P^\pi(x_3,b_1)
      +\frac{1}{36}r_\pi\Bigl((2-x_2-x_3)E_q
      -(x_2-x_3)k_1^3\Bigr)\phi_\sigma^{\prime K}(x_2,b_2)\\ &\times
    \phi_\sigma^{\prime\pi}(x_3,b_1)\biggr]h_d(x_1,1-x_2,1-x_3,b_1,b_2)
    \exp[-S_B(t_d^2)-S_K(t_d^2)-S_\pi(t_d^2)|_{b_3\rightarrow b_1}]\biggr\},\nonumber  
  \end{split}
\end{equation}

\begin{equation} \label{eq:mek}
  \begin{split}
    \mathcal{M}_{eK}=&2\pi^2f_Bf_Kf_\pi m_B^2\frac{C_F}{N_c}
      \int k_{1\perp}dk_{1\perp}\int_{x_1^d}^{x_1^u}dx_1\int_0^1dx_2dx_3
      \int_0^\infty b_1db_1b_3db_3
      \left(\frac{1}{2}m_B+\frac{|\mathbf{k}_{1\perp}|^2}{2x_1^2m_B}\right)
      K(\vec{k}_1)\\ &\times (E_Q+m_Q)
    J_0(k_{1\perp}b_1)\phi_\pi(x_3,b_3)
      \biggl\{\alpha_s(t_{dK}^1)
        \biggl[-x_3(E_q-k_1^3)\phi_K(x_2,b_1)+r_K(1-x_2)(E_q+k_1^3)\phi_P^K(x_2,b_1)
    \\ &+\frac{1}{6}r_K(1-x_2)(E_q+k_1^3)\phi_\sigma^{\prime K}(x_2,b_1)\biggr]
      h_d(x_1,x_3,1-x_2,b_1,b_3))\exp[-S_B(t_{dK}^1)-S_K(t_{dK}^1)|_{b_2\rightarrow b_1}\\ &-S_\pi(t_{dK}^1)]
    +\alpha_s(t_{dK}^2)
      \biggl[\Bigl((2-x_2-x_3)E_q-(x_2-x_3)k_1^3\Bigr)\phi_K(x_2,b_1)
      -r_K(1-x_2)(E_q-k_1^3)\phi_P^K(x_2,b_1) \\
    &+\frac{1}{6}r_K(1-x_2) (E_q-k_1^3)\phi_\sigma^{\prime K}(x_2,b_1)\biggr]
      h_d(x_1,1-x_3,1-x_2,b_1,b_3)
      \exp[-S_B(t_{dK}^2)\\ &-S_K(t_{dK}^2)|_{b_2\rightarrow b_1}-S_\pi(t_{dK}^2)]\biggr\}, 
  \end{split}
\end{equation}
\begin{equation} \label{eq:mekp}
  \begin{split}
    \mathcal{M}_{eK}^P=&2\pi^2f_Bf_Kf_\pi m_B^2\frac{C_F}{N_c}
      \int k_{1\perp}dk_{1\perp}\int_{x_1^d}^{x_1^u}dx_1\int_0^1dx_2dx_3
      \int_0^\infty b_1db_1b_3db_3
      \left(\frac{1}{2}m_B+\frac{|\mathbf{k}_{1\perp}|^2}{2x_1^2m_B}\right)
      K(\vec{k}_1)\\ &\times(E_Q+m_Q)
    J_0(k_{1\perp}b_1)\phi_\pi(x_3,b_3)
      \biggl\{\alpha_s(t_{dK}^1)
        \biggl[\Bigl((1-x_2+x_3)E_q+(1-x_2-x_3)k_1^3\Bigr)\phi_K(x_2,b_1)
          -r_K\\ &\times(1-x_2)(E_q-k_1^3)\phi_P^K(x_2,b_1)
    +\frac{1}{6}r_K(1-x_2)(E_q-k_1^3)\phi_\sigma^{\prime K}(x_2,b_1)\biggr]
      h_d(x_1,x_3,1-x_2,b_1,b_3) \\
    &\times\exp[-S_B(t_{dK}^1)-S_K(t_{dK}^1)|_{b_2\rightarrow b_1}-S_\pi(t_{dK}^1)]
    +\alpha_s(t_{dK}^2)
      \biggl[-(1-x_3)(E_q-k_1^3)\phi_K(x_2,b_1)
      +r_K(1-x_2)\\ &\times(E_q+k_1^3)\phi_P^K(x_2,b_1)
    +\frac{1}{6}r_K(1-x_2)(E_q+k_1^3)\phi_\sigma^{\prime K}(x_2,b_1)\biggr]
      h_d(x_1,1-x_3,1-x_2,b_1,b_3) \\ &\times
      \exp[-S_B(t_{dK}^2)-S_K(t_{dK}^2)|_{b_2\rightarrow b_1}-S_\pi(t_{dK}^2)]\biggr\}. 
  \end{split}
\end{equation}
The amplitudes of the nonfactorizable annihilation diagrams (e) and (f) are
\begin{equation} \label{eq:ma}
  \begin{split}
    \mathcal{M}_{a}=&2\pi^2f_Bf_Kf_\pi m_B^2\frac{C_F}{N_c}\int k_{1\perp}dk_{1\perp}
      \int_{x_1^d}^{x_1^u}dx_1\int_0^1dx_2dx_3
      \int_0^\infty b_1db_1b_2db_2
      \left(\frac{1}{2}m_B+\frac{|\mathbf{k}_{1\perp}|^2}{2x_1^2m_B}\right)
      K(\vec{k}_1)(E_Q+m_Q) \\
    &\times J_0(k_{1\perp}b_1)
      \biggl\{\alpha_s(t_f^1)
        \biggl[-x_3(E_q-k_1^3)\phi_K(x_2,b_2)\phi_\pi(x_3,b_2)
          -r_Kr_\pi\Bigl((1-x_2+x_3)E_q
          +(1-x_2-x_3)k_1^3\Bigr) \\ & \times\phi_P^K(x_2,b_2)\phi_P^\pi(x_3,b_2)
    +\frac{1}{6}r_Kr_\pi\Bigl((1-x_2-x_3)E_q
      +(1-x_2+x_3)k_1^3\Bigr)\phi_P^K(x_2,b_2)\phi_\sigma^{\prime\pi}(x_3,b_2)
    -\frac{1}{6}r_Kr_\pi\\ &\times\Bigl((1-x_2-x_3)E_q
      +(1-x_2+x_3)k_1^3\Bigr)\phi_\sigma^{\prime K}(x_2,b_2)\phi_P^\pi(x_3,b_2)
      +\frac{1}{36}r_Kr_\pi\Bigl((1-x_2+x_3)E_q
    +(1-x_2\\ &-x_3)k_1^3\Bigr)
      \phi_\sigma^{\prime K}(x_2,b_2)\phi_\sigma^{\prime\pi}(x_3,b_2)\biggr]
      h_f^1(1-x_2,x_3,b_1,b_2)
      \exp[-S_B(t_f^1)-S_K(t_f^1)-S_\pi(t_f^1)|_{b_3\rightarrow b_2}] \\
    &+\alpha_s(t_f^2)
      \biggl[(1-x_2)(E_q+k_1^3)\phi_K(x_2,b_2)\phi_\pi(x_3,b_2)
      +r_Kr_\pi\Bigl((3-x_2+x_3)E_q
      +(1-x_2-x_3)k_1^3\Bigr) \\ & \times\phi_P^K(x_2,b_2)\phi_P^\pi(x_3,b_2)
    +\frac{1}{6}r_Kr_\pi\Bigl((1-x_2-x_3)E_q
      -(1+x_2-x_3)k_1^3\Bigr)\phi_P^K(x_2,b_2)\phi_\sigma^{\prime\pi}(x_3,b_2) \\
    &-\frac{1}{6}r_Kr_\pi\Bigl((1-x_2-x_3)E_q
      +(3-x_2+x_3)k_1^3\Bigr)\phi_\sigma^{\prime K}(x_2,b_2)\phi_P^\pi(x_3,b_2)
      +\frac{1}{36}r_Kr_\pi\Bigl((1+x_2-x_3)E_q  \\
      &-(1-x_2-x_3)k_1^3\Bigr)\phi_\sigma^{\prime K}(x_2,b_2)\phi_\sigma^{\prime\pi}(x_3,b_2)\biggr]
      h_f^2(1-x_2,x_3,b_1,b_2)
      \exp[-S_B(t_f^2)-S_K(t_f^2)-S_\pi(t_f^2)|_{b_3\rightarrow b_2}]\biggr\}, 
  \end{split}
\end{equation}
\begin{equation} \label{eq:map}
  \begin{split}
    \mathcal{M}_{a}^P=&2\pi^2f_Bf_Kf_\pi m_B^2\frac{C_F}{N_c}\int k_{1\perp}dk_{1\perp}
      \int_{x_1^d}^{x_1^u}dx_1\int_0^1dx_2dx_3
      \int_0^\infty b_1db_1b_2db_2
      \left(\frac{1}{2}m_B+\frac{|\mathbf{k}_{1\perp}|^2}{2x_1^2m_B}\right)
      K(\vec{k}_1)(E_Q+m_Q) \\
    &\times J_0(k_{1\perp}b_1)
      \biggl\{\alpha_s(t_f^1)
        \biggl[r_\pi x_3(E_q+k_1^3)\phi_K(x_2,b_2)\phi_P^\pi(x_3,b_2)
        +\frac{1}{6}r_\pi x_3(E_q+k_1^3)\phi_K(x_2,b_2)\phi_\sigma^{\prime\pi}(x_3,b_2)
    -r_K\\ &\times(1-x_2)(E_q-k_1^3)\phi_P^K(x_2,b_2)\phi_\pi(x_3,b_2)
      +\frac{1}{6}r_K(1-x_2)(E_q-k_1^3)\phi_\sigma^{\prime K}(x_2,b_2)\phi_\pi(x_3,b_2) \biggr] \\
    &\times h_f^1(1-x_2,x_3,b_1,b_2)\exp[-S_B(t_f^1)-S_K(t_f^1)-S_\pi(t_f^1)|_{b_3\rightarrow b_2}]
    +\alpha_s(t_f^2)
      \biggl[r_\pi\Bigl((2-x_3)E_q+x_3k_1^3\Bigr)\\ &\times\phi_K(x_2,b_2)\phi_P^\pi(x_3,b_2)
      +\frac{1}{6}r_\pi\Bigl((2-x_3)E_q+x_3k_1^3\Bigr)\phi_K(x_2,b_2)\phi_\sigma^{\prime\pi}(x_3,b_2)
    -r_K\Bigl((1+x_2)E_q-(1-x_2)k_1^3\Bigr)\\
    &\times\phi_P^K(x_2,b_2)\phi_\pi(x_3,b_2)
      +\frac{1}{6}r_K\Bigl((1+x_2)E_q-(1-x_2)k_1^3\Bigr)
        \phi_\sigma^{\prime K}(x_2,b_2)\phi_\pi(x_3,b_2) \biggr]
    h_f^2(1-x_2,x_3,b_1,b_2)\\ & \times\exp[-S_B(t_f^2)-S_K(t_f^2)-S_\pi(t_f^2)|_{b_3\rightarrow b_2}]\biggr\}. 
  \end{split}
\end{equation}
\end{widetext}
In the Eqs.~\eqref{eq:fe}--\eqref{eq:map}, we have defined
\begin{equation}
  r_{K,\pi}=\mu_{K,\pi}/m_B=m_{K,\pi}^2/[(m_{s,u}+m_d)m_B],
\end{equation}
$C_F=4/3$ and $N_c=3$ are color factors. The function $h$'s are derived from the Fourier transformation of hard amplitudes. $S_{B,K,\pi}(t)$ are Sudakov factors and $S_t(x)$ is threshold factor. They are all given in Appendix~\ref{ap:factors}. The expressions of $b$-space wave functions $\phi_{M}(x,b)$, $\phi_P^M(x,b)$ and
$\phi_\sigma^M(x,b)$ ($M=K,\; \pi$) can be found in Appendix~\ref{ap:kaonwf}. Particularly, the factor $\chi_B$ in Eq.~\eqref{eq:fap} is defined by
\begin{equation} \label{eq:chib}
  \begin{split}
    \chi_B=&\pi\int dk_{1\perp}k_{1\perp}\int_{x_1^d}^{x_1^u}dx_1
    \left(\frac{1}{2}m_B+\frac{|\mathbf{k}_{1\perp}|^2}{2x_1^2m_B}\right) \\
    &\times K(\vec{k}_1)\left[(E_q+m_q)(E_Q+m_Q)+|\vec{k}_1|^2\right],
  \end{split}
\end{equation}
which comes from the $B$ to vacuum matrix element with the $(S-P)$ operator inserted.

In order to decrease the high-order corrections, the renormalization scales $t$ are taken as the
largest virtualities in the decay amplitudes
\begin{equation}
  \begin{split}
    t_e^1=&\max(\sqrt{1-x_3}m_B, 1/b_1, 1/b_3),\\
    t_e^2=&\max(\sqrt{x_1}m_B, 1/b_1, 1/b_3),
  \end{split}
\end{equation}
\begin{equation}
  \begin{split}
    t_{eK}^1=&\max(\sqrt{1-x_2}m_B, 1/b_1, 1/b_2),\\
    t_{eK}^2=&\max(\sqrt{x_1}m_B, 1/b_1, 1/b_2),
  \end{split}
\end{equation}
\begin{equation}
  \begin{split}
    t_a^1=&\max(\sqrt{x_3}m_B, 1/b_2, 1/b_3),\\
    t_a^2=&\max(\sqrt{1-x_2}m_B, 1/b_2, 1/b_3),
  \end{split}
\end{equation}
and
\begin{equation}
  \begin{split}
    t_d^1=\max(&\sqrt{x_1(1-x_3)}m_B, \\
      &\quad\sqrt{x_2(1-x_3)}m_B, 1/b_1, 1/b_2),\\
    t_d^2=\max(&\sqrt{x_1(1-x_3)}m_B, \\
      &\quad \sqrt{(1-x_2)(1-x_3)}m_B, 1/b_1, 1/b_2),
  \end{split}
\end{equation}

\begin{equation}
  \begin{split}
    t_{dK}^1=\max(&\sqrt{x_1(1-x_2)}m_B, \\
      &\quad\sqrt{(1-x_2)x_3}m_B, 1/b_1, 1/b_3),\\
    t_{dK}^2=\max(&\sqrt{x_1(1-x_2)}m_B, \\
      &\quad\sqrt{(1-x_2)(1-x_3)}m_B, 1/b_1, 1/b_3),
  \end{split}
\end{equation}
\begin{equation}
  \begin{split}
    t_f^1=\max(&\sqrt{(1-x_2)x_3}m_B, 1/b_1, 1/b_2),\\
    t_f^2=\max(&\sqrt{(1-x_2)x_3}m_B, \\
      &\quad\sqrt{1-x_2+x_2x_3}m_B, 1/b_1, 1/b_2).
  \end{split}
\end{equation}

In the language of the above matrix elements for different diagrams, i.e., Eqs. \eqref{eq:fe}--\eqref{eq:map},
the decay amplitudes for $B\rightarrow K\pi$ decays can be written as:
\begin{widetext}
\begin{equation} \label{eq:amp1}
  \begin{split}
    \mathcal{M}&({B}^-\rightarrow \bar{K}^0\pi^-)\\
    =&-V_t\Bigl(\frac{C_3}{N_c}+C_4-\frac{1}{2}\frac{C_9}{N_c}-\frac{1}{2}C_{10}\Bigr)f_KF_e
      -V_t\Bigl(\frac{C_5}{N_c}+C_6-\frac{1}{2}\frac{C_7}{N_c}-\frac{1}{2}C_8\Bigr)f_K {F}_{e}^P
      -\frac{V_t}{N_c}\Bigl(C_3-\frac{1}{2}C_9\Bigr)\mathcal{M}_{e}\\
  \end{split}
\end{equation}
\begin{equation} 
  \begin{split}
     &-\frac{V_t}{N_c}\Bigl(C_5-\frac{1}{2}C_7\Bigr)\mathcal{M}_{e}^P
      +\frac{1}{N_c}\biggl[V_uC_1-V_t(C_3+C_9)\biggr]\mathcal{M}_{a}
      -\frac{V_t}{N_c}(C_5+C_7)\mathcal{M}_{a}^P\\
    &+\biggl[V_u\Bigl(\frac{C_1}{N_c}+C_2\Bigr)
      -V_t\Bigl(\frac{C_3}{N_c}+C_4+\frac{C_9}{N_c}+C_{10}\Bigr)\biggr]f_BF_{a}
      -V_t\Bigl(\frac{C_5}{N_c}+C_6+\frac{C_7}{N_c}+C_8\Bigr)f_BF_{a}^P,\nonumber
  \end{split}
\end{equation}
\begin{equation} \label{eq:amp2}
  \begin{split}
    \sqrt{2}\mathcal{M}&({B}^-\rightarrow {K}^-\pi^0)\\
    =&\biggl[V_u\Bigl(\frac{C_1}{N_c}+C_2\Bigr)
      -V_t\Bigl(\frac{C_3}{N_c}+C_4+\frac{C_9}{N_c}+C_{10}\Bigr)\biggr]f_KF_{e}
      -V_t\Bigl(\frac{C_5}{N_c}+C_6+\frac{C_7}{N_c}+C_8\Bigr)f_KF_{e}^P\\
    &+\biggl[V_u\Bigl(C_1+\frac{C_2}{N_c}\Bigr)
      -\frac{3V_t}{2}\Bigl(-C_7-\frac{C_8}{N_c}+C_9+\frac{C_{10}}{N_c}\Bigr)\biggr]f_\pi F_{eK}
      +\frac{1}{N_c}\biggl[V_uC_1-V_t(C_3+C_9)\biggr]\mathcal{M}_{e}\\
    &-\frac{V_t}{N_c}(C_5+C_7)\mathcal{M}_{e}^P
      +\frac{1}{N_c}\Bigl(V_u{C_2}-\frac{3V_t}{2}C_{10}\Bigr)\mathcal{M}_{eK}
      -\frac{V_t}{N_c}\frac{3}{2}C_8\mathcal{M}_{eK}^P
      +\frac{1}{N_c}\biggl[V_uC_1-V_t(C_3+C_9)\biggr]\mathcal{M}_{a}\\
    &-\frac{V_t}{N_c}(C_5+C_7)\mathcal{M}_{a}^P
      +\biggl[V_u\Bigl(\frac{C_1}{N_c}+C_2\Bigr)
      -V_t\Bigl(\frac{C_3}{N_c}+C_4+\frac{C_9}{N_c}+C_{10}\Bigr)\biggr]f_BF_{a}\\
    &-V_t\Bigl(\frac{C_5}{N_c}+C_6+\frac{C_7}{N_c}+C_8\Bigr)f_BF_{a}^P,\\
  \end{split}
\end{equation}
\begin{equation} \label{eq:amp3}
  \begin{split}
    \mathcal{M}&(\bar{B}^0\rightarrow K^-\pi^+)\\
    =&\biggl[V_u\Bigl(\frac{C_1}{N_c}+C_2\Bigr)
      -V_t\Bigl(\frac{C_3}{N_c}+C_4+\frac{C_9}{N_c}+C_{10}\Bigr)\biggr]f_KF_{e}
      -V_t\Bigl(\frac{C_5}{N_c}+C_6+\frac{C_7}{N_c}+C_8\Bigr)f_KF_{e}^P\\
    &+\frac{1}{N_c}\biggl[V_uC_1-V_t(C_3+C_9)\biggr]\mathcal{M}_{e}
      -\frac{V_t}{N_c}(C_5+C_7)\mathcal{M}_{e}^P
      -\frac{V_t}{N_c}\Bigl(C_3-\frac{1}{2}C_9\Bigr)\mathcal{M}_{a}
      -\frac{V_t}{N_c}\Bigl(C_5-\frac{1}{2}C_7\Bigr)\mathcal{M}_{a}^P\\
    &-V_t\Bigl(\frac{C_3}{N_c}+C_4-\frac{1}{2}\frac{C_9}{N_c}-\frac{1}{2}C_{10}\Bigr)
      f_B {F}_{a}
      -V_t\Bigl(\frac{C_5}{N_c}+C_6-\frac{1}{2}\frac{C_7}{N_c}-\frac{1}{2}C_8\Bigr)
      f_B {F}_{a}^P,\\
  \end{split}
\end{equation}
\begin{equation} \label{eq:amp4}
  \begin{split}
    -\sqrt{2}\mathcal{M}&(\bar{B}^0\rightarrow \bar{K}^0\pi^0)\\
    =&-V_t\Bigl(\frac{C_3}{N_c}+C_4-\frac{1}{2}\frac{C_9}{N_c}-\frac{1}{2}C_{10}\Bigr)f_KF_e
      -V_t\Bigl(\frac{C_5}{N_c}+C_6-\frac{1}{2}\frac{C_7}{N_c}-\frac{1}{2}C_8\Bigr)f_K {F}_{e}^P
    -\biggl[V_u\Bigl(C_1+\frac{C_2}{N_c}\Bigr)\\ &
      -\frac{3V_t}{2}\Bigl(-C_7-\frac{C_8}{N_c}+C_9+\frac{C_{10}}{N_c}\Bigr)\biggr]f_\pi F_{eK}
      -\frac{V_t}{N_c}\Bigl(C_3-\frac{1}{2}C_9\Bigr)\mathcal{M}_{e}
      -\frac{V_t}{N_c}\Bigl(C_5-\frac{1}{2}C_7\Bigr)\mathcal{M}_{e}^P\\
    &-\frac{1}{N_c}\Bigl(V_u{C_2}-\frac{3V_t}{2}C_{10}\Bigr)\mathcal{M}_{eK}
      +\frac{V_t}{N_c}\frac{3}{2}C_8\mathcal{M}_{eK}^P
      -\frac{V_t}{N_c}\Bigl(C_3-\frac{1}{2}C_9\Bigr)\mathcal{M}_{a}
      -\frac{V_t}{N_c}\Bigl(C_5-\frac{1}{2}C_7\Bigr)\mathcal{M}_{a}^P\\
    &-V_t\Bigl(\frac{C_3}{N_c}+C_4-\frac{1}{2}\frac{C_9}{N_c}-\frac{1}{2}C_{10}\Bigr)f_BF_a
      -V_t\Bigl(\frac{C_5}{N_c}+C_6-\frac{1}{2}\frac{C_7}{N_c}-\frac{1}{2}C_8\Bigr)f_B{F}_{a}^P,\\
  \end{split}
\end{equation}
with $N_c=3$.
\end{widetext}

The decay width is calculated by
\begin{equation}
    \Gamma(B\rightarrow K\pi)=\frac{G_F^2m_B^3}{128\pi}|\mathcal{M}(B\rightarrow K\pi)|^2.
\end{equation}
And the expressions of branching ratios and direct \textit{CP} violations are
\begin{equation}
  Br(B\rightarrow K\pi)=\Gamma(B\rightarrow K\pi)/\Gamma_B,
\end{equation}
\begin{equation}
  \begin{split}
    &A_{CP}(B^0(B^+)\rightarrow K\pi)\\
    &\quad =\frac{\Gamma(\bar{B}^0(B^-)\rightarrow K\pi)-\Gamma(B^0(B^+)\rightarrow K\pi)}
      {\Gamma(\bar{B}^0(B^-)\rightarrow K\pi)+\Gamma(B^0(B^+)\rightarrow K\pi)}.
  \end{split}
\end{equation}

The Sudakov factor suppresses nonperturbative contributions and makes PQCD approach applicable
\cite{li1996perturbative,li1996pqcd}. However, the suppression effect of Sudakv factor depends on the end-point behavior of wave functions. With the $B$ meson wave function obtained by solving the bound-state equation in relativistic potential model \cite{yang2012wave,liu2014spectrum,liu2015spectrum,sun2017decay,sun2019wave},
we find the suppression of Sudakov factor to soft contribution is not strong enough. To restore the reliability of perturbative calculation, we introduce the momentum cutoff and soft form factor under a critical scale $\mu_c=1.0~\textup{GeV}$, which corresponds to the strong coupling constant $\alpha_s/\pi =0.165$. The contributions lower than the cutoff scale $\mu_c $ are removed and replaced by the relevant soft form factors.
The effect of soft form factor will be investigated in Sec.~\ref{sec:softff}.

\section{The next-to-leading order corrections} \label{sec:nlo}
In order to improve the results, the most important next-to-leading-order (NLO) corrections
to the decay amplitudes from the vertex corrections, the quark loops, and the magnetic penguins are included. These contributions have been considered in PQCD approach in Ref.~\cite{li2005resolution}.
It turns out that the NLO corrections affect the amplitudes by changing the Wilson coefficients.
For simplicity, we define the combinations of Wilson coefficients
\begin{equation}
  \begin{split}
    a_1(\mu)=&C_2(\mu)+\frac{C_1(\mu)}{N_c},\\
    a_2(\mu)=&C_1(\mu)+\frac{C_2(\mu)}{N_c},\\
    a_i(\mu)=&C_i(\mu)+\frac{C_{i\pm 1}(\mu)}{N_c},\\
  \end{split}
\end{equation}
with $i=3-10$. When $i$ is odd (even), the plus (minus) sign is taken.

\subsection{Vertex Corrections}
At first, we consider the vertex corrections.
Since the corrections of nonfactorizable diagrams are negligible
and the annihilation diagrams themselves do not contribute much to the amplitudes,
we concentrate on the vertex corrections of the factorizable emission diagrams,
i.e., diagrams (a) and (b) in Fig.~\ref{fig:diagrams}.
The vertex corrections modify the Wilson coefficients as
\cite{beneke1999qcd, beneke2000qcd, beneke2001qcd, li2005resolution}
\begin{equation}
  \begin{split}
    a_1(\mu)&\rightarrow a_1(\mu)+\frac{\alpha_s(\mu)}{4\pi}C_F\frac{C_1(\mu)}{N_c}V_1(M),\\
    a_2(\mu)&\rightarrow a_2(\mu)+\frac{\alpha_s(\mu)}{4\pi}C_F\frac{C_2(\mu)}{N_c}V_2(M),\\
    a_i(\mu)&\rightarrow a_i(\mu)+\frac{\alpha_s(\mu)}{4\pi}C_F\frac{C_{i\pm 1}(\mu)}{N_c}V_i(M),\\
  \end{split}
\end{equation}
with $i=3-10$, and $M$ represents the meson emitted from the weak vertex.
For $V_{1,4,6,8,10}$, $M$ is kaon and for $V_{2,3,5,7,9}$, $M$ is pion.
In the naive dimensional regularization (NDR) scheme $V_i(M)$ are given by \cite{beneke1999qcd}
\begin{equation}
  \begin{split}
    &V_i(M)=\\
      &\left\{
        \begin{array}{l}
          12\ln(m_b/\mu)-18+\int_0^1dx\phi_M^A(x)g(x),\vspace{2mm} \\
            \hspace{4cm} \textup{for}~i=1-4,9,10 \vspace{2mm}\\
          -12\ln(m_b/\mu)+6-\int_0^1dx\phi_M^A(x)g(1-x), \vspace{2mm}\\
            \hspace{4cm} \textup{for}~i=5,7 \vspace{2mm}\\
          -6+\int_0^1dx\phi_M^P(x)h(x), \quad\  \textup{for}~i=6,8 \\
        \end{array}
        \right.
  \end{split}
\end{equation}
with $\phi_M^A(x)$ and $\phi_M^P(x)$ are the twist-2 and twist-3 meson distribution amplitudes, and $x$ the parton momentum fraction. The functions $g(x)$ and $h(x)$ are defined by
\begin{equation}
  \begin{split}
    g(x)=&3\left(\frac{1-2x}{1-x}\ln x-i\pi\right)+\biggl[2\textup{Li}_2(x)-\ln^2x \\
      &+\frac{2\ln x}{1-x}-(3+2i\pi)\ln x-(x\longleftrightarrow 1-x)\biggr], \\
  \end{split}
\end{equation}
\begin{equation}
  h(x)=2\textup{Li}_2(x)-\ln^2x-(1+2i\pi)\ln x-(x\longleftrightarrow 1-x).
\end{equation}

\subsection{Quark Loops}
For the $B\rightarrow K\pi$ decays, the effective Hamiltonian of the virtual quark loops are given by \cite{li2005resolution}
\begin{equation}
  \begin{split}
    H_{\textup{eff}}=&-\sum_{q=u,c,t}\sum_{q^\prime}\frac{G_F}{\sqrt{2}}V_{qb}V_{qs}^*
      \frac{\alpha_s(\mu)}{2\pi}C^{(q)}(\mu,l^2)\\
    &\quad\times(\bar{s}\gamma_\rho(1-\gamma_5)T^ab)(\bar{q}^\prime\gamma^\rho T^aq^\prime),\\
    \end{split}
\end{equation}
where $l^2$ is the invariant mass of the gluon.
The functions $C^{(q)}(\mu,l^2)$ are
\begin{equation} \label{eq:cquc}
  C^{(q)}(\mu,l^2)=\left[G^{(q)}(\mu,l^2)-\dfrac{2}{3}\right]C_2(\mu),
\end{equation}
for $q=u,c$, and
\begin{equation} \label{eq:cqt}
  \begin{split}
    &C^{(q)}(\mu,l^2)\\
    &\quad=\left[G^{(s)}(\mu,l^2)-\dfrac{2}{3}\right]C_3(\mu)\\
    &\qquad+\sum\limits_{q^{\prime\prime}=u,d,s,c}G^{(q^{\prime\prime})}(\mu,l^2)
        \left[C_4(\mu)+C_6(\mu)\right],\\
  \end{split}
\end{equation}
for $q=t$.
The function $G^{(q)}(\mu,l^2)$ shown in Eq.~\eqref{eq:cquc} and \eqref{eq:cqt} for the loop of
the quark $q$ is given by
\begin{equation}
  \begin{split}
    &G^{(q)}(\mu,l^2)\\
    &\quad=-4\int_0^1dxx(1-x)\ln\frac{m_q^2-x(1-x)l^2-i\epsilon}{\mu^2},
  \end{split}
\end{equation}
where $m_q$ is the quark mass.

Because the topological structure of quark loops is similar with the penguin diagrams,
its effect can be absorbed into the Wilson coefficients $a_4$ and $a_6$ by
\begin{equation}
  \begin{split}
    a_{4,6}(\mu)&\rightarrow a_{4.6}(\mu)+\frac{\alpha_s(\mu)}{9\pi}
      \sum\limits_{q=u,c,t}\frac{V_{qb}V_{qs}^*}{V_{tb}V_{ts}^*}C^{(q)}(\mu,\langle l^2\rangle),\\
  \end{split}
\end{equation}
where $\langle l^2\rangle$ is the mean virtual gluon-momentum squared in the decay process. In our numerical calculations of $B\rightarrow K\pi$ decays, $\langle l^2\rangle=m_b^2/4$ is taken as an average gluon momentum squared, which is an reasonable value in $B$ decays.

\subsection{Magnetic Penguins}
Then, we investigate the correction from the magnetic penguin.
The effective Hamiltonian of the magnetic penguin contains the $b\rightarrow sg$ transition
\begin{equation}
  H_{\textup{eff}}=-\frac{G_F}{\sqrt{2}}V_{tb}V_{ts}^*C_{8g}O_{8g},
\end{equation}
and the magnetic penguin operator is
\begin{equation}
  O_{8g}=\frac{g_s}{8\pi^2}m_b\bar{s}_i\sigma_{\mu\nu}(1+\gamma_5)T_{ij}^aG^{a\mu\nu}b_j,
\end{equation}
with the color indices $i$ and $j$.
Considering the similar topological structure of magnetic penguin and quark loop,
we can also absorb the contribution of magnetic penguin operator into the Wilson coefficients
\cite{li2005resolution}
\begin{equation}
  \begin{split}
    &a_{4,6}\rightarrow a_{4,6}-\frac{\alpha_s(\mu)}{9\pi}
      \frac{2m_B}{\sqrt{\langle l^2\rangle}}C_{8g}^{\textup{eff}}(\mu),\\
  \end{split}
\end{equation}
with the effective Wilson coefficient $C_{8g}^{\textup{eff}}=C_{8g}+C_5$ \cite{buchalla1996weak}.

\section{The Contribution of the soft $BK$, $B\pi$ Transition and $K\pi$ Production Form Factors}
\label{sec:softff}
With Eqs.~\eqref{eq:fe}-\eqref{eq:map}, we calculate the eight topological diagrams shown in Fig.~\ref{fig:diagrams} numerically, and find that soft contributions associated with diagrams (a), (b), (g) and (h) are not negligible. For the diagrams (a), (b), (g) and (h), there are more than $40\%$ contributions in the range of $\alpha_s/\pi>0.2$ which are not in good perturbative region. In contrast, the soft contributions are only a few percent in diagrams (c), (d), (e) and (f). In order to improve the reliability of perturbative calculation,
we introduce the momentum cutoff and soft form factors for scales lower than the critical scale  $\mu_c=1.0~\textup{GeV}$. That means we treat the contributions with the scale $\mu<\mu_c$ as nonperturbative quantity and replace them by phenomenological soft form factors. The soft contributions are absorbed into two kinds of soft form factors, the $BK$, $B\pi$ transition form factors and the $K\pi$ production form factor.

With the soft transition form factors, we can express the $B\rightarrow K$ and $B\rightarrow\pi$
transition form factors as
\begin{equation} \label{eq:fbkpi}
  \begin{split}
    &F_0^{BK}=h_0^{BK}+\xi^{BK},\\
    &F_0^{B\pi}=h_0^{B\pi}+\xi^{B\pi},\\
  \end{split}
\end{equation}
where $h_0^{BK}$ and $h_0^{B\pi}$ represent the hard contributions which can be evaluated perturbatively
in PQCD approach, and $\xi^{BK}$ and $\xi^{B\pi}$ are soft $BK$, $B\pi$ transition form factors.
In result, the amplitudes in Eqs.~\eqref{eq:amp1}-\eqref{eq:amp4} should be modified by
\begin{equation}
  \begin{split}
    \mathcal{M}&(B\rightarrow K\pi)\rightarrow \mathcal{M}(B\rightarrow K\pi) \\
      &-2C_\pi(\mu_c)V_{\textup{CKM}}f_\pi\xi^{BK}-4r_\pi C_\pi^\prime(\mu_c)V_{\textup{CKM}}f_\pi\xi^{BK} \\
      &-2C_K(\mu_c)V_{\textup{CKM}}f_K\xi^{B\pi}-4r_KC_K^\prime(\mu_c)V_{\textup{CKM}}f_K\xi^{B\pi}, \\
  \end{split}
\end{equation}
where $V_{\textup{CKM}}$ represents the relevant CKM matrix elements, $r_{K,\pi}$ the parameters related with mesons kaon and pion which have been defined in Sec.~\ref{sec:perturlo}, $C_\pi$ and $C_\pi^\prime$ the appropriate combinations of Wilson coefficients for the diagrams with pion emitted out and with $(V-A)(V-A)$ and $(S+P)(S-P)$ operators inserted respectively, $C_K$ and $C_K^\prime$ are Wilson coefficients for kaon emitted diagrams. These Wilson coefficients are taken at the cutoff scale $\mu_c$, which is the critical separation scale of hard and soft contributions.

Furthermore, we also introduce the soft $K\pi$ production form factor $\xi^{K\pi}$ to absorb the soft
contribution in the factorizable annihilation diagrams (g) and (h) in Fig.~\ref{fig:diagrams}.
The $K\pi$ form factor can be defined by the matrix element of scalar current
\begin{equation}
  \begin{split}
    \langle& K\pi|S|0\rangle=-\frac{1}{2}\sqrt{\mu_K\mu_\pi}F_+^{K\pi},
  \end{split}
\end{equation}
where $\mu_{K,\pi}=m_{K,\pi}^2/(m_{s,u}+m_d)$.
Considering the soft part, the $K\pi$ production form factor can be written as
\begin{equation}
  F_+^{K\pi}\rightarrow h^{K\pi}+\xi^{K\pi},
\end{equation}
where $h^{K\pi}$ is the hard part that can be calculated perturbatively according to the factorizable annihilation
diagrams, and $\xi^{K\pi}$ the soft part of the $K\pi$ production form factor. With the soft $K\pi$ production form factor $\xi^{K\pi}$, the amplitudes are changed as
\begin{equation}
  \begin{split}
    \mathcal{M}&(B\rightarrow K\pi)\rightarrow \\
      &\mathcal{M}(B\rightarrow K\pi)-2\sqrt{r_Kr_\pi}C(\mu_c)V_{\textup{CKM}}\chi_Bf_B\xi^{K\pi},
  \end{split}
\end{equation}
where $f_B$ is the decay constant of $B$ meson, $\chi_B$ the parameter defined in Eq.~\eqref{eq:chib},
and $r_{K,\pi}=\mu_{K,\pi}/m_B$.

\section{The Contribution of Color-Octet Matrix Element} \label{sec:co}
To improve the consistency between the theoretical calculation and experiment data,
we take into account the contribution from the hadronic matrix element of color-octet operators.
By considering the relation for the generators of the color SU(3) group
\begin{equation}\label{generatorTT}
  T^a_{ik}T^a_{jl}=-\frac{1}{2N_c}\delta_{ik}\delta_{jl}+\frac{1}{2}\delta_{il}\delta_{jk},
\end{equation}
we can decompose the four-quark operators with any Dirac spinor structure into color-singlet and color-octet operators
\begin{equation}
  \begin{split} \label{quarkcolor}
    &(\bar{q}_{1i}q_{2j})(\bar{q}_{3j}q_{4i}) \\
      &\quad=\frac{1}{N_c}(\bar{q}_{1i}q_{2i})(\bar{q}_{3j}q_{4j})
      +2(\bar{q}_{1i}T^a_{ik}q_{2k})(\bar{q}_{3j}T^a_{jl}q_{4l}), \\
  \end{split}
\end{equation}
where the first and second terms correspond to color-singlet and -octet operators, respectively, and the specific Dirac spinor structure is omitted for simplicity.

To make clear how to obtain the contribution of color-octet operators, let us take the contribution of the tree operators $O_1^u$ and $O_2^u$ to the decay amplitude of $\bar{B}^0\to K^- \pi^+$ at leading order as an example. The contribution of $O_1^u$ and $O_2^u$ is
\begin{eqnarray}\label{example1}
A&& = \langle K^- \pi^+|C_1 O_1^u + C_2 O_2^u|\bar{B}^0\rangle \nonumber \\
  && = C_1 \langle K^-\pi^+ |(\bar{s}_i\gamma^\mu(1-\gamma_5)u_j)
  (\bar{u}_{j}\gamma_\mu(1-\gamma_5)b_i)|\bar{B}^0\rangle\nonumber \\
  &&\quad + C_2 \langle K^-\pi^+ |(\bar{s}_i\gamma^\mu(1-\gamma_5)u_i)
  (\bar{u}_{j}\gamma_\mu(1-\gamma_5)b_j)|\bar{B}^0\rangle \nonumber \\
\end{eqnarray}
where the quarks that form the first meson in the final state are always moved to the first current in the above equation. If the quarks that form the same meson are not in the same current in the original Hamiltonian, Fierz transformation should be performed. The quark pair in the first current in the second term of Eq. (\ref{example1}) can form $K^-$ directly at leading order in QCD, but the first term is not the case. So Eqs. (\ref{generatorTT}) and (\ref{quarkcolor}) should be used for the currents in the first term, then Eq. (\ref{example1}) becomes
\begin{eqnarray} \label{colorA}
&&A=\nonumber \\
  &&(\frac{C_1}{N_c} + C_2) \langle K^-\pi^+ |(\bar{s}_i\gamma^\mu(1-\gamma_5)u_i)
  (\bar{u}_{j}\gamma_\mu(1-\gamma_5)b_j)|\bar{B}^0\rangle\nonumber \\
  &&+ 2 C_1 \langle K^-\pi^+ |(\bar{s}\gamma^\mu(1-\gamma_5)T^a u)
  (\bar{u}\gamma_\mu(1-\gamma_5)T^a b)|\bar{B}^0\rangle\nonumber \\
\end{eqnarray}
The last term is the color-octet contribution, which is usually dropped previously as a hadronic matrix element of long-distance quantity, because the mesons should be in color-singlet states. Note that no hard gluon exchange between color-octet quark-antiquark pairs should be further considered in the color-octet matrix element in Eq. (\ref{colorA}) because such matrix element is a quantity dominated by long-distance dynamics. Hard gluons can only be transferred at short distance, which is higher order corrections in perturbative QCD as shown by the non-factorizable diagrams in Fig.~\ref{fig:diagrams}, where the quark-antiquark pairs in the final state are color-singlet. The results of the short-distance calculation given in Eqs.~\eqref{eq:fe}-\eqref{eq:map} do not include the contribution of the color-octet matrix element. Only contribution of color-singlet quark pairs at hadronic level should be contained in these equations.

The color-octet hadronic matrix element are defined as a non-perturbative quantity at hadronic level, where only soft-gluon can exchange between the color-octet quark-antiquark pairs.

The above procedure can be done in a different way, i.e., we can treat the Hamiltonian by using Eqs. (\ref{generatorTT}) and (\ref{quarkcolor}) at first, then use it to the decay at hadronic level. The tree-level Hamiltonian operator is
\begin{eqnarray}
&& C_1 O_1^u + C_2 O_2^u \nonumber\\
  && = C_1 (\bar{s}_i\gamma^\mu(1-\gamma_5)u_j)
  (\bar{u}_{j}\gamma_\mu(1-\gamma_5)b_i)\nonumber\\
  &&\quad + C_2 (\bar{s}_i\gamma^\mu(1-\gamma_5)u_i)
  (\bar{u}_{j}\gamma_\mu(1-\gamma_5)b_j)
\end{eqnarray}
Using Eqs. (\ref{generatorTT}) and (\ref{quarkcolor}) to decompose the color non-singlet operator in the above equation into singlet and octet operators, then it becomes
\begin{eqnarray}
  && C_1 O_1^u + C_2 O_2^u \nonumber\\
  &&=(\frac{C_1}{N_c} + C_2) (\bar{s}_i\gamma^\mu(1-\gamma_5)u_i)
  (\bar{u}_{j}\gamma_\mu(1-\gamma_5)b_j)\nonumber\\
  &&\quad + 2 C_1 (\bar{s}\gamma^\mu(1-\gamma_5)T^a u)
  (\bar{u}\gamma_\mu(1-\gamma_5)T^a b)  \label{octet-Hamiltonian}
\end{eqnarray}
The effective Hamiltonian operator in the above form is only different from the original one by different operator bases. If we use Eq. (\ref{octet-Hamiltonian}) to $\bar{B}^0\to K^- \pi^+$ decay again, we can obtain exactly the same result as what is given in Eq. (\ref{colorA}). So the color-octet contribution of hadronic matrix element actually stems from the relevant color-octet operator contained in the original Hamiltonian. It is not difficult for this discussion to be extended to penguin operators in the effective Hamiltonian and operators with which Fierz transformation needs to be performed.  Therefore, the treatment for color-octet contribution will not cause confusion or double counting.

 For the $B\rightarrow K\pi$ decays, at the leading order approximation, the hadronic matrix element of color-singlet operator can be written as
 \begin{equation} \label{eq:t0}
   \begin{split}
     T_{K\pi}^0&=\langle K\pi|(\bar{s}\gamma^\mu(1-\gamma_5)q)(\bar{q}\gamma_\mu(1-\gamma_5)b)|\bar{B}\rangle \\
       &\approx -if_Km_B^2F_0^{B\pi}(0),\\
     T_{K\pi}^{SP0}&=\langle K\pi|(\bar{s}(1+\gamma_5)q)(\bar{q}(1-\gamma_5)b)|\bar{B}\rangle \\
       &\approx -if_Kr_Km_B^2F_0^{B\pi}(0),\\
   \end{split}
 \end{equation}
 where $q\in \{u,d\}$, $S$ and $P$ stand for scalar and pseudoscalar currents. Whereas, up to now there is no reliable way to estimate the value for the color-octet hadronic matrix elements, which are defined as
 \begin{equation} \label{eq:t8}
   \begin{split}
     T_{K\pi}^8&=\langle K\pi|(\bar{s}T^a\gamma^\mu(1-\gamma_5)q)(\bar{q}T^a\gamma_\mu(1-\gamma_5)b)|\bar{B}\rangle, \\
     T_{K\pi}^{SP8}&=\langle K\pi|(\bar{s}T^a(1+\gamma_5)q)(\bar{q}T^a(1-\gamma_5)b)|\bar{B}\rangle. \\
   \end{split}
 \end{equation}
 In Eqs.~\eqref{eq:t0} and \eqref{eq:t8}, the quarks in the first current make up the first meson in the final
 state and the quarks in the second current involve the meson in the initial state and the second meson in the final state. Actually there should also be $T_{\pi K}^{(SP)0}$ matrix elements in the calculation, but considering $f_KF_0^{B\pi}(0)\approx f_\pi F_0^{BK}(0)$, the difference between $T_{K\pi}^{(SP)0}$ and $T_{\pi K}^{(SP)0}$ can be safely neglected.

The color-octet hadronic matrix elements are usually dropped previously in the literature, because the hadronic states should be color-singlet. However, color-octet quark-antiquark states can change to be color-singlet states by exchanging soft gluons at distance of hadronic scale. Therefore, the contribution of the color-octet hadronic matrix element may not be zero from the theoretical point of view. In this work, we take the color-octet contributions into consideration. With the approximation in Eq. (\ref{eq:t0}), one can define the color-octet parameter $\delta_8$ and $\delta_8^{SP}$ through the color-octet hadronic matrix elements in the following way
 \begin{equation} \label{TKpi-delta8}
   \begin{split}
     T_{K\pi}^8&=-if_Km_B^2F_0^{B\pi}(0)\delta_8,\\
     T_{K\pi}^{SP8}&=-if_Kr_Km_B^2F_0^{B\pi}(0)\delta_8^{SP},\\
   \end{split}  
 \end{equation}
 so that $\delta_8$ and $\delta_8^{SP}$ can be viewed as a measure of how large the color-octet matrix element compared to color-singlet matrix element.

In the treatment of the color-octet contribution in the above, no hard-gluon exchanging effect between the out-emitting quark-antiquark system and the remaining quark system is considered. This corresponds to only considering the color-octet contribution from the factorizable diagrams (a) and (b) in Fig.~\ref{fig:diagrams}. Actually we can consider the color-octet contributions of both the factorizable and non-factorizable diagrams in Fig.~\ref{fig:diagrams} in a consistent way by separating the color-singlet and octet components in the calculation of the diagrams in Fig.~\ref{fig:diagrams} by a procedure described in the following. Here we consider the case that the initial $B$ meson is in the color-singlet state and only the light quark-antiquark pairs that form the final mesons of pion and kaon can be in the color-octet state, and treat the momentum-distribution of the quarks in the color-octet system the same as that in pion and kaon, i.e., we define the color-octet quark-antiquark pair in a state similar to pion and kaon. Then the wave functions of the color-octet quark-antiquark pair are defined in a similar way to Eq. (\ref{pion-wave} ) by
\begin{equation} \label{pion-wave-8}
  \begin{split}
    &\langle\pi^8,K^8(p)|\bar{q}_\delta(x)T^a q_\gamma^\prime(0)|0\rangle \\
    &\quad=\int dud^2k_{q\perp}\Phi_{\gamma\delta}^{\pi,K}
      \exp\left[i(up\cdot x-x_\perp\cdot k_{q\perp})\right], \\
  \end{split}
\end{equation}
where $\langle\pi^8,K^8(p)|$ denote color-octet state of quark-antiquark system that are finally transferred to pion and kaon, and the spinor wave function $\Phi_{\gamma\delta}^{\pi,K}$ is taken to be the same thing as that of the color-singlet case in Eq. (\ref{pion-spinor-wave} ). For the case of kaon, the wave functions of pion should be replaced by that of kaon.

The effect of the color-octet quark-antiquark system transferring to color-singlet state by exchanging soft gluons is considered by introducing a multiplying parameter $Y^8$. In the numerical treatment, we find two parameters are needed to explain the experimental data of $B\to K\pi$ decays, $Y^8_F$ and $Y^8_M$, which are relevant to the factorizable and non-factorizable diagrams in Fig.~\ref{fig:diagrams}, respectively. The relation between the parameters $\delta_8$ and $Y^8_F$ can be easily obtained by considering the color-octet contribution of the factorizable diagrams (a) and (b) of Fig.~\ref{fig:diagrams} in these two different way. In the following analysis, we treat the color-octet effect in terms of the parameters $Y^8_F$ and $Y^8_M$.

\begin{figure}[b]
  \includegraphics[width=0.45\textwidth]{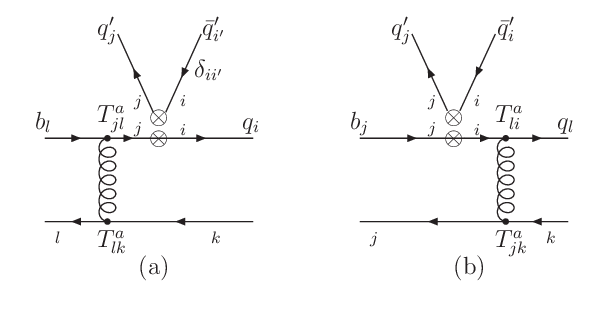}
  \caption{\label{fig:diagrams-ab} Factorizable diagrams with operator insertion of $(\bar{q}_ib_j)(\bar{q}^\prime_j q^\prime_i)$, where the explicit current type such as $(V\pm A)$ or $(S\pm P)$ are omitted. The quark-antiquark pairs in the final state are in non-singlet color states. The symbols $i$, $i^\prime$, $j$, $k$ and $l$ are all color indices.}
\end{figure}

The color-octet contribution according to each diagram of Fig.~\ref{fig:diagrams} can be calculated by considering the quark-antiquark pairs in the final states in color non-singlet, which includes both the color-singlet and -octet components. The color-octet component can be separated from color-singlet one by analyzing the color factors in each diagram. To show the analyzing procedure clearly, we draw the diagrams (a) and (b) of Fig.~\ref{fig:diagrams} again in Fig.~\ref{fig:diagrams-ab} with the color indices $i$, $i^\prime$, $j$, $k$ and $l$ for each quark shown explicitly in it. As an example, let us consider the case with the operator  $(\bar{q}_i b_j)(\bar{q}^\prime_j q^\prime_i)$ inserted in the diagrams, where $i$ and $j$ are the color indices, and the current can be either $(V\pm A)$ or $(S\pm P)$ etc., which are not shown explicitly. The color indices of the quark pairs in the final sate are $(i^\prime, ~j)$, $(i,~k)$ in Fig.~\ref{fig:diagrams-ab} (a), and $(i,~j)$, $(k,~l)$ in Fig.~\ref{fig:diagrams-ab} (b), respectively. All the color indices are summed in the calculation because they should be summed for both the color-singlet and -octet states for the quark pairs in the initial and final states.

In the calculation of the hard amplitude according to Fig.~\ref{fig:diagrams-ab} (a) and (b), the momentum convolutions are just the same as that for the color-singlet cases given in Eqs.~\eqref{eq:fe}-\eqref{eq:map}, the only difference is for the color factors. So we need only to analyze the color factors in the case that the quark pairs in the final state are in color non-singlet state. For Fig.~\ref{fig:diagrams-ab} (a), the color factor becomes

\begin{equation}\label{eq:colorfactora}
  \begin{split}
    \sum_{ijkl} T_{jl}^a T_{lk}^a &= \sum_{ijk} C_F \delta_{jk}
      = \sum_{ijki^\prime} C_F \delta_{jk} \delta_{ii^\prime} \\
    & = \sum_{ijki^\prime} C_F \left(\dfrac{1}{N_c} \delta_{ji^\prime} \delta_{ik} + 2 T_{ji^\prime}^a T_{ik}^a\right),
  \end{split}
\end{equation}
and for Fig.~\ref{fig:diagrams-ab} (b) the color factor is
\begin{equation}\label{eq:colorfactorb}
  \begin{split}
    &\sum_{ijkl} T_{li}^a T_{jk}^a
    = \sum_{ijkl} \left[-\dfrac{1}{2N_c} \delta_{li} \delta_{jk} + \dfrac{1}{2} \delta_{lk} \delta_{ji} \right] \\
  &\quad = \sum_{ijkl} \left[-\dfrac{1}{2N_c} \left(\dfrac{1}{N_c} \delta_{lk} \delta_{ji}
      + 2 T_{lk}^b T_{ji}^b\right) + \dfrac{1}{2} \delta_{lk} \delta_{ji} \right]\\
    &\quad = \sum_{ijkl} \left(\dfrac{C_F}{N_c} \delta_{lk} \delta_{ji}
      - \dfrac{1}{N_c} T_{lk}^b T_{ji}^b \right),
  \end{split}
\end{equation}
where the first terms with two delta factors correspond to the color-singlet contribution, while the second terms the color-octet contribution. The first terms give exactly $F_e$ and $F_e^P$ given in Eqs. (\ref{eq:fe}) and (\ref{eq:fep}) but with an extra color-suppression factor $1/N_c$. The second terms in Eqs. (\ref{eq:colorfactora}) and (\ref{eq:colorfactorb}) will give the contribution for the quark pairs in the final state being in color-octet state. After the parameters $Y^8_F$ and $Y^8_M$ are introduced that describe the effect of changing the quark pair of color-octet into color-singlet state by exchanging soft gluons, the color-octet contribution of each diagram can be obtained. The contribution of Fig.~\ref{fig:diagrams-ab} (a) and (b) for $(V-A)(V-A)$ and $(S+P)(S-P)$ operators are
\begin{equation}\label{eq:fetofe8}
 Y^8_F F_e^{(P)8},
\end{equation}
where
\begin{equation}\label{eq:fe8}
  F_e^{(P)8} \equiv 2N_c^2 F_e^{(P)a} - \dfrac{N_c}{C_F} F_e^{(P)b}.
\end{equation}
The contribution of the color-octet component in the other diagrams in Fig.~\ref{fig:diagrams} with operator insertion of all kinds of currents are
\begin{equation}
Y^8_M\mathcal{M}_e^{(P)8},\;
Y^8_M\mathcal{M}_e^{(P)\prime 8},\;Y^8_M\mathcal{M}_a^{(P)8},\; Y^8_F F_a^{(P)8},
\end{equation}
where
\begin{equation}\label{fa8}
  \begin{split}
    &\mathcal{M}_e^{(P)8} \equiv 2 N_c^2 \mathcal{M}_e^{(P)c} - \dfrac{N_c}{C_F} \mathcal{M}_e^{(P)d}, \\
    &\mathcal{M}_e^{(P)\prime 8} \equiv \dfrac{N_c^2}{C_F} \mathcal{M}_e^{(P)}, \
      \mathcal{M}_a^{(P)8} \equiv -\dfrac{N_c}{C_F} \mathcal{M}_a^{(P)}\\
      &      F_a^{(P)8} \equiv -\dfrac{N_c^2}{C_F} F_a^{(P)}.
  \end{split}
\end{equation}
Here the parameter $Y^8_{F(M)}$ is the parameters that describes the effect of color-octet state transferring into color-singlet by exchanging soft gluons for the factorizable (non-factorizable) diagrams. The quantities $F_e^{(P)a}$, $F_e^{(P)b}$, $\mathcal{M}_e^{(P)c}$, and $\mathcal{M}_e^{(P)d}$ are the convolution functions for the diagrams (a) $\sim$ (d) in Fig.~\ref{fig:diagrams}, which will not be given apparently here for simplicity, while $\mathcal{M}_e^{(P)}$, $\mathcal{M}_a^{(P)}$  and $F_a^{(P)}$ have been given in Eqs. \eqref{eq:fa}-\eqref{eq:map}. 

From Eq. (\ref{eq:fetofe8}), it is known that the color-octet contribution to the amplitude corresponding to Fig.~\ref{fig:diagrams-ab} (a) and (b) with operator insertion of $(V-A)(V-A)$ current is
\begin{equation}\label{expression-a}
 Y^8_F F_e^{8},
\end{equation}
where a common factor $i\frac{G_F}{\sqrt{2}}\frac{m_B^2}{2}$, $f_K$, the CKM matrix element and Wilson coefficients are omitted. The same quantity can be expressed in terms of the color-octet hadronic matrix element $T_{K\pi}^8$ in Eq. (\ref{TKpi-delta8}) too, which is
\begin{equation}\label{expression-b}
(i\frac{m_B^2}{2}f_K)^{-1}2T_{K\pi}^8=-4F_0^{B\pi}(0)\delta_8.
\end{equation}
The quantities of Eqs. (\ref{expression-a}) and (\ref{expression-b}) can be identified as the same, so the relation of $Y^8_F$ and $\delta_8$ can be obtained as
\begin{equation}\label{reletion-Y8-delta8}
Y^8_F F_e^{8}=-4F_0^{B\pi}(0)\delta_8.
\end{equation}

Including the color-octet contributions, the amplitudes in Eqs. \eqref{eq:amp1}-\eqref{eq:amp4} are
modified by

\begin{equation} \label{eq:y8amp1}
 \begin{split}
     \mathcal{M}&({B}^-\rightarrow \bar{K}^0\pi^-) \rightarrow \mathcal{M}({B}^-\rightarrow \bar{K}^0\pi^-)\\
     & +\biggl\{-V_t \Bigl(C_3 - \frac{1}{2}C_9\Bigr) f_K F_e^8
       -V_t \Bigl(C_5 - \frac{1}{2}C_7\Bigr) f_K F_e^{P8}\\
     &+\biggl[V_u\Bigl(\frac{C_1}{N_c}+C_2\Bigr)
        -V_t\Bigl(\frac{C_3}{N_c}+C_4+\frac{C_9}{N_c}+C_{10}\Bigr)\biggr] \\
 & \cdot f_BF_a^8-V_t\Bigl(\frac{C_5}{N_c}+C_6+\frac{C_7}{N_c}+C_8\Bigr)f_BF_a^{P8}\biggr\}Y_F^8\\
      & +\biggl\{-V_t \Bigl(C_3-\frac{1}{2}C_9\Bigr)\mathcal{M}_e^8
       -V_t\Bigl(C_4-\frac{1}{2}C_{10}\Bigr)\mathcal{M}_e^{\prime 8} \\
     & -V_t\Bigl(C_5-\frac{1}{2}C_7\Bigr)\mathcal{M}_e^{P8}
       -V_t\Bigl(C_6-\frac{1}{2}C_8\Bigr)\mathcal{M}_e^{P\prime 8}\\
      & +\biggl[V_uC_1-V_t(C_3+C_9)\biggr]\mathcal{M}_a^8
       -V_t(C_5+C_7)\mathcal{M}_a^{P8} \biggr\} \\
       &\cdot Y_M^8,
   \end{split}
 \end{equation}
   \begin{equation} \label{eq:y8amp2}
   \begin{split}
     \sqrt{2}\mathcal{M}&({B}^-\rightarrow {K}^-\pi^0) \rightarrow
       \sqrt{2}\mathcal{M}({B}^-\rightarrow {K}^-\pi^0) \\
     & +\biggl\{\biggl[V_u C_1 -V_t \Bigl(C_3 + C_9\Bigr)\biggr] f_K F_e^8
       -V_t\Bigl(C_5 \\
       &+ C_7\Bigr) f_K F_e^{P8}+\biggl[V_u C_2 -\frac{3V_t}{2}\Bigl(-C_8 + C_{10}\Bigr)\biggr] \\
        & \cdot f_\pi F_{eK}^8+\biggl\{\biggl[V_u C_1 - V_t (C_3 + C_9)\biggr] \mathcal{M}_e^8
       +\biggl[V_u C_2\\
       & - V_t (C_4+ C_{10})\biggr] \mathcal{M}_e^{\prime 8} -V_t (C_5 + C_7) \mathcal{M}_e^{P8}\\
       &-V_t (C_6 + C_8) \mathcal{M}_e^{P\prime 8} +\Bigl(V_u C_2 - \frac{3V_t}{2} C_{10}\Bigr) \mathcal{M}_{eK}^8\\
       &+\Bigl(V_u C_1 - \frac{3V_t}{2} C_9\Bigr) \mathcal{M}_{eK}^{\prime 8}
       -\frac{3V_t}{2} C_8 \mathcal{M}_{eK}^{P8}\\
       &-\frac{3V_t}{2} C_7 \mathcal{M}_{eK}^{P\prime 8}
      +\biggl[V_u C_1 - V_t (C_3 + C_9)\biggr] \mathcal{M}_a^8\\
      &-V_t (C_5 + C_7) \mathcal{M}_a^{P8}\biggr\} Y_M^8,
 \end{split}
 \end{equation}
\begin{equation} \label{eq:y8amp3}
   \begin{split}
     \mathcal{M}&(\bar{B}^0\rightarrow K^-\pi^+) \rightarrow
       \mathcal{M}(\bar{B}^0\rightarrow K^-\pi^+) \\
     & +\biggl\{\biggl[V_u C_1 - V_t \Bigl(C_3 + C_9\Bigr)\biggr] f_K F_e^8
       -V_t \Bigl(C_5 + C_7\Bigr)\\
       &\cdot f_K F_e^{P8}
       -V_t \Bigl(\frac{C_3}{N_c} + C_4 - \frac{1}{2}\frac{C_9}{N_c} - \frac{1}{2}C_{10}\Bigr) f_B F_a^8 \\
     & -V_t \Bigl(\frac{C_5}{N_c} + C_6 - \frac{1}{2}\frac{C_7}{N_c} - \frac{1}{2}C_8\Bigr) f_B F_a^{P8}
       \biggr\} Y_F^8\\
     & +\biggl\{\biggl[V_u C_1 - V_t (C_3 + C_9)\biggr] \mathcal{M}_e^8
       +\biggl[V_u C_2 - V_t (C_4 \\
     &+ C_{10})\biggr] \mathcal{M}_e^{\prime 8}-V_t (C_5 + C_7) \mathcal{M}_e^{P8}
       -V_t (C_6 + C_8) \mathcal{M}_e^{P\prime 8}\\
     &  -V_t \Bigl(C_3 - \frac{1}{2} C_9\Bigr) \mathcal{M}_a^8
       -V_t \Bigl(C_5 - \frac{1}{2} C_7\Bigr) \mathcal{M}_a^{P8} \biggr\} Y_M^8 ,
   \end{split}
 \end{equation}

 \begin{widetext}
  \begin{equation} \label{eq:y8amp4}
    \begin{split}
      -\sqrt{2}\mathcal{M}&(\bar{B}^0\rightarrow \bar{K}^0\pi^0) \rightarrow
        -\sqrt{2}\mathcal{M}(\bar{B}^0\rightarrow \bar{K}^0\pi^0) \\
      & +\biggl\{-V_t \Bigl(C_3 - \frac{1}{2} C_9\Bigr) f_K F_e^8
        -V_t \Bigl(C_5 - \frac{1}{2} C_7\Bigr) f_K F_e^{P8}
        -\biggl[V_u C_2 - \frac{3V_t}{2} \Bigl(-C_8 + C_{10}\Bigr)\biggr] f_\pi F_{eK}^8 \\
      & -V_t \Bigl(\frac{C_3}{N_c} + C_4 - \frac{1}{2} \frac{C_9}{N_c} - \frac{1}{2}C_{10}\Bigr) f_B F_a^8
        -V_t \Bigl(\frac{C_5}{N_c} + C_6 - \frac{1}{2} \frac{C_7}{N_c} - \frac{1}{2}C_8\Bigr) f_B F_a^{P8}
        \biggr\} Y_F^8\\
      & +\biggl\{-V_t \Bigl(C_3 - \frac{1}{2} C_9\Bigr) \mathcal{M}_e^8
        -V_t \Bigl(C_4 - \frac{1}{2} C_{10}\Bigr) \mathcal{M}_e^{\prime 8}
        -V_t \Bigl(C_5 - \frac{1}{2} C_7\Bigr) \mathcal{M}_e^{P8}
        -V_t \Bigl(C_6 - \frac{1}{2} C_8\Bigr) \mathcal{M}_e^{P\prime 8}\\
      & -\Bigl(V_u C_2 - \frac{3V_t}{2} C_{10}\Bigr) \mathcal{M}_{eK}^8
        -\Bigl(V_u C_1 - \frac{3V_t}{2} C_9\Bigr) \mathcal{M}_{eK}^{\prime 8}
        +\frac{3V_t}{2} C_8 \mathcal{M}_{eK}^{P8}
        +\frac{3V_t}{2} C_7 \mathcal{M}_{eK}^{P\prime 8}\\
      & -V_t \Bigl(C_3 - \frac{1}{2} C_9\Bigr) \mathcal{M}_a^8
        -V_t \Bigl(C_5 - \frac{1}{2} C_7\Bigr) \mathcal{M}_a^{P8}\biggr\} Y_M^8.
    \end{split}
  \end{equation}
\end{widetext}
The value of color-octet parameters $Y_{F,M}^8$ can be determined from experiment data. It is reasonable that the magnitude of them should not be too large.

\section{Numerical Result and Discussion} \label{sec:result}
In addition to the parameters in $B$ meson, pion and kaon wave functions, there are several other numerical parameters in the calculation, which are the soft $BK$, $B\pi$ transition form factors $\xi^{BK}$ and $\xi^{B\pi}$,
the soft $K\pi$ production form factors $\xi^{K\pi}$, and the color-octet matrix element parameters $Y_F^8$ and $Y_M^8$. 

The hard part of the $BK$ and $B\pi$ transition form factors $h_0^{BK}$ and $h_0^{B\pi}$ can be calculated directly in perturbative QCD with the momentum cutoff in scale $\mu>\mu_c= 1.0~\textup{GeV}$, which is relevant to the diagrams (a) and (b) in Fig.~\ref{fig:diagrams}. The results of hard transition form factors are
\begin{equation}\label{hffactor}
  \begin{split}
    &h_0^{BK}=0.29\pm 0.02,\\
    &h_0^{B\pi}=0.23\pm 0.01.\\
  \end{split}
\end{equation}
For the total $BK$ and $B\pi$ transition form factors, we take 
\begin{equation}\label{ffactor}
  \begin{split}
    &F_0^{BK}=0.33\pm 0.04,\\
    &F_0^{B\pi}=0.27\pm 0.02.
  \end{split}
\end{equation}
For the form factor $F_0^{BK}$ at large recoil limit, several calculations of Lattice QCD (LQCD) can be found in the literature \cite{PBD2023,Bailey2016}, which are shown in Table \ref{tab:fbk}. The values in the first tow columns are from calculations of LQCD, and the third one from light-cone sum rule (LCSR) \cite{ball2005new} for comparison. The value of $F_0^{BK}$ given in Eq. (\ref{ffactor}) is the average of the two results of LQCD. Here only two effective digits are kept for accuracy consistence in this work.

\begin{table*}
  \caption{\label{tab:fbk}The value of $B\to K$ form factor $F_+^{B K}(q^2=0)=F_0^{B K}(q^2=0)$.}
  \renewcommand\arraystretch{1.5}
  \begin{ruledtabular}
    \begin{tabular}{cccc}
      & LQCD-HQPCD \cite{PBD2023}
      & LQCD-FNAL \cite{Bailey2016}
      & LCSR \cite{ball2005new} \\
      \hline
      $F_+^{B K}(q^2=0)$ & $0.332(12)$ & $0.335(36)$ & $0.331\pm 0.041$
    \end{tabular}
  \end{ruledtabular}
\end{table*}

The value of $F_0^{B\pi}$ is taken by considering both the result of LQCD in \cite{UKQCD1998}, which is $F_{0(\mathrm{LQCD})}^{B\pi }=0.27\pm 0.11$, and the experimental data on the differential branching fraction of $B\to \pi\ell\nu$ decay around $q^2=0$ \cite{sibidanov2013study}, here $q^2$ is the invariant mass of the lepton pair. We only take 0.02 as the variation of the theoretical input in our numerical calculation without taking the large error given in \cite{UKQCD1998}, because most part of the form factor within the large error band will make the semileptonic decay branching ratio exceed the experimental upper limit. The value of $F_0^{B\pi}$ given in Eq.(\ref{ffactor}) is consistent with the experimental data in Ref. \cite{sibidanov2013study} when it is used to calculate the semileptonic decay branching ratio with the CKM matrix elements given in PDG \cite{Zyla:2020zbs}.

Using Eqs.~\eqref{eq:fbkpi}, \eqref{hffactor} and \eqref{ffactor}, we can obtain the following result for the soft part of the transition form factors
\begin{equation}\label{eq:xibkandxibpi}
  \begin{split}
   &\xi^{BK}=0.04\pm 0.02,\\
   &\xi^{B\pi}=0.04\pm 0.01.\\
  \end{split}
\end{equation}

As for the color-octet parameters $Y_F^8$ and $Y_M^8$, there is not a systematical way to evaluate the values of them up to now. We will treat them as free parameters, and determine them by experimental data. The nonperturbative parameters $Y_F^8$, $Y_M^8$ and $\xi^{K\pi}$ can be written in the following form
\begin{equation} \label{parameter-n}
  \begin{split}
    \xi^{K\pi}& = d_1 \exp(i\phi_1),\\
    Y_F^8& = d_2 \exp(i\phi_2),\\
    Y_M^8& = d_3 \exp(i\phi_3),\\
  \end{split}
\end{equation}
where $d_{1,2,3}$ and $\phi_{1,2,3}$ are the magnitudes and phases of these parameters, respectively. For $B\rightarrow K\pi$ decays, there are data for branching ratios and \textit{CP} violations of four decay modes, which can be used to determine these parameters. By fitting to the experimental data, we find the values of parameters are
\begin{equation}\label{eq:dphi}
  \begin{split}
    &d_1 = 0.10_{-0.01}^{+0.01},\quad \phi_1 = (0.47_{-0.03}^{+0.05})\pi,\\
    &d_2 = 0.12_{-0.02}^{+0.01},\quad \phi_2 = (1.00_{-0.04}^{+0.03})\pi,\\
    &d_3 = 0.05_{-0.03}^{+0.03},\quad \phi_3 = (0.93_{-0.14}^{+0.17})\pi,\\
  \end{split}
\end{equation}
where the uncertainties come from the constraint of experimental data. With the value of $Y_F^8$ in the above two equations and using Eq. (\ref{reletion-Y8-delta8}), we can get
\begin{equation}
\delta_8=0.27^{+0.02}_{-0.05}\exp(1.00^{+0.03}_{-0.04}i\pi),
\end{equation}
which shows that the absolute value of the color-octet hadronic matrix element is indeed not large compared with the case of color-singlet one.

Using the parameters given in Eqs. (\ref{parameter-n}) and (\ref{eq:dphi}), the predictions of $B\rightarrow K\pi$ branching ratios and \textit{CP} violations are

\begin{equation}
  \begin{split}
    &B({B}^+\rightarrow K^0\pi^+) = 24.3_{-4.7-2.3}^{+4.5+2.4} \times 10^{-6},\\
    &B({B}^+\rightarrow K^+\pi^0) = 12.6_{-2.5-1.0}^{+2.3+1.1} \times 10^{-6},\\
    &B({B}^0\rightarrow K^+\pi^-) = 20.0_{-3.7-1.2}^{+3.4+1.3} \times 10^{-6},\\
    &B({B}^0\rightarrow K^0\pi^0) =  9.4_{-1.8-0.8}^{+1.7+0.8} \times 10^{-6},\\
    &A_{CP}(B^+\rightarrow K^0\pi^+) =  0.012_{-0.001-0.001}^{+0.001+0.001},\\
    &A_{CP}(B^+\rightarrow K^+\pi^0) =  0.041_{-0.028-0.015}^{+0.034+0.012},\\
    &A_{CP}(B^0\rightarrow K^+\pi^-) = -0.084_{-0.038-0.049}^{+0.035+0.044},\\
    &A_{CP}(B^0\rightarrow K^0\pi^0) = -0.112_{-0.055-0.040}^{+0.050+0.036},
  \end{split}
\end{equation}
where the first uncertainty comes from the uncertainties of the nonperturbative parameters in Eqs. \eqref{eq:xibkandxibpi} and \eqref{eq:dphi}, and the second one from the uncertainties of the parameters in the meson wave functions.

\begin{table*}
  \caption{\label{tab:res}
    $B\rightarrow K\pi$ branching ratios and \textit{CP} violations.}
  \renewcommand\arraystretch{1.5}
  \begin{ruledtabular}
    \begin{tabular}{cccccccccc}
      Mode
      & $\mathrm{LO}$
      & $\mathrm{LO_{NLOWC}}$
      & $\mathrm{NLO}$
      & $+\xi^{BK(\pi)}$ 
      & $+\xi^{K\pi}$ 
      & $+Y_F^8$ 
      & $+Y_M^8$ 
      & $+\xi^{BK(\pi)}+\xi^{K\pi}+Y_{F,M}^8$
      & {Data}\cite{Zyla:2020zbs} \\
      \hline
      $B({B}^+\rightarrow K^0\pi^+)\times 10^{-6}$& $8.5 $ & $13.4$ & $13.8$ & $20.8$ & $14.0$ & $22.9$ & $11.2$ & $24.3_{-4.7-2.3}^{+4.5+2.4}$         &$23.7\pm 0.8$ \\
      $B({B}^+\rightarrow K^+\pi^0)\times 10^{-6}$& $6.0 $ & $9.0 $ & $8.4 $ & $12.0$ & $7.6 $ & $12.5$ & $6.8 $ & $12.6_{-2.5-1.0}^{+2.3+1.1}$         &$12.9\pm 0.5$ \\
      $B({B}^0\rightarrow K^+\pi^-)\times 10^{-6}$& $8.8 $ & $13.7$ & $13.2$ & $18.8$ & $11.5$ & $21.4$ & $11.5$ & $20.0_{-3.7-1.2}^{+3.4+1.3}$         &$19.6\pm 0.5$ \\
      $B({B}^0\rightarrow K^0\pi^0)\times 10^{-6}$& $2.9 $ & $4.9 $ & $5.2 $ & $7.9 $ & $5.2 $ & $9.6 $ & $4.7 $ & $ 9.4_{-1.8-0.8}^{+1.7+0.8}$         &$9.9 \pm 0.5$ \\
      $A_{CP}(B^+\rightarrow K^0\pi^+)$           &$-0.006$&$-0.004$&$ 0.010$&$ 0.013$&$ 0.013$&$ 0.006$&$ 0.010$&$ 0.012_{-0.001-0.001}^{+0.001+0.001}$&$-0.017\pm 0.016$ \\
      $A_{CP}(B^+\rightarrow K^+\pi^0)$           &$-0.185$&$-0.153$&$-0.039$&$-0.001$&$ 0.073$&$-0.032$&$-0.003$&$ 0.041_{-0.028-0.015}^{+0.034+0.012}$&$ 0.037\pm 0.021$ \\
      $A_{CP}(B^0\rightarrow K^+\pi^-)$           &$-0.239$&$-0.175$&$-0.107$&$-0.063$&$ 0.025$&$-0.195$&$-0.126$&$-0.084_{-0.038-0.049}^{+0.035+0.044}$&$-0.083\pm 0.004$ \\
      $A_{CP}(B^0\rightarrow K^0\pi^0)$           &$ 0.004$&$ 0.018$&$-0.036$&$-0.040$&$-0.048$&$-0.147$&$-0.094$&$-0.112_{-0.055-0.040}^{+0.050+0.036}$&$ 0.00 \pm 0.13 $ \\
    \end{tabular}
  \end{ruledtabular}
\end{table*}

The contributions of each theoretical component and the total results for the branching ratios and $CP$ violations are listed in Table~\ref{tab:res}. The experimental data are also presented in the last column for comparison. In Table~\ref{tab:res}, the column ``LO" means the hard contribution of leading-order in QCD with leading-order Wilson coefficients being used, ``$\mathrm{LO_{NLOWC}}$" the LO results with NLO Wilson coefficients being used, ``NLO" the hard contribution up to next-to-leading order in QCD, ``$+\xi^{BK(\pi)}$" the contribution of NLO plus the contribution of the soft transition form factor $\xi^{BK}$ and $\xi^{B\pi}$, ``$+\xi^{K\pi}$" the contribution of NLO plus the contribution of soft production form factor of $K\pi$,
``$+Y_{F(M)}^8$" the contribution of NLO plus  color-octet matrix element in factorizable (non-factorizable) diagrams, and ``$+\xi^{BK(\pi)}+\xi^{K\pi}+Y_{F,M}^8$" the total contribution of NLO$+\xi^{BK(\pi)}+\xi^{K\pi}+Y_{F,M}^8$. The difference of the columns labeled by LO and $\mathrm{LO_{NLOWC}}$ is only caused by the difference of LO and NLO Wilson coefficients. The branching ratios increase much because of the penguin enhancement effect of NLO Wilson coefficients, which is consistent with what has been found in \cite{li2005resolution}. The difference of the column ``NLO" and ``$\mathrm{LO_{NLOWC}}$" reflects the effects of NLO contributions of the vertex correction, quark-loop and magnetic penguin in QCD expansion. The branching ratios are only slightly changed by these NLO contributions, which shows the efficiency of the perturbative expansion in the modified PQCD approach. Table~\ref{tab:res} also shows that, the branching ratios of only hard calculation up to next-to-leading order are much smaller than experimental data.
By introducing the soft $BK$ and $B\pi$ transition form factors $\xi^{BK}$ and $\xi^{B\pi}$,
the branching ratios are increased greatly,
but \textit{CP} violations still deviates significantly from experimental data.
The soft $K\pi$ production form factor $\xi^{K\pi}$ has a small impact on the branching ratio,
but it causes a change in the sign of the \textit{CP} violation in $B^+\rightarrow K^+\pi^0$.
The contribution of the color-octet matrix element in factorizable diagrams $Y_F^8$ significantly increases
the branching ratios of all four decay channels,
and the color-octet matrix element in nonfactorizable diagrams $Y_M^8$ increases the \textit{CP} violation
of $B^+\rightarrow K^+\pi^0$ channel.
They are essential for explaining the experimental data.
By comparing the last two columns of Table~\ref{tab:res}, we can find that the theoretical results of all the branching ratios and most of the \textit{CP} violations for $B\rightarrow K\pi$ decays are consistent with the experiment data. \textit{CP} violation for $B^+\rightarrow K^0\pi^+$ is very close to the data considering both the experimental and theoretical uncertainties.

A few comments would like to be given here.

1) As soft quantities, it is possible that $\xi^{M_1M_2}$ and $Y_{F,M}^8(M_1 M_2)$, where $M_1M_2$ may be any possible final states such as $\pi\pi$ and $K\pi$ etc., depend on the mesonic final states. The values $\xi^{\pi\pi}$ and $\delta_8^{\pi\pi}$ obtained in the study of $B\to\pi\pi$ decays in Ref. \cite{luyang2022} is shown to be different from the values of $\xi^{K\pi}$ and $\delta_8^{K\pi}$ obtained in this work.
One possible explanation is that the soft production form factors may depend on the center-of-mass energy of the meson pair, the inner relative moving state between them and the wave functions of the mesons. If any one of the dependence is sensitive, then the soft form factors can be apparently different between $\pi\pi$ and $K\pi$ final states. So does the hadronic color-octet matrix element $Y_{F(M)}^8$ used in this work. To make the method used in this work predictive, one way is to study these soft quantities independently by a completely nonperturbative method, or to try to find relations between the soft quantities with different final states in a phenomenological way, which is beyond the scope of this paper. But we hope this can be achieved in the near future.

2) The soft transition form factors defined in this work depend on the critical scale that separates the soft and hard interaction. In principle, such critical scale can not be fixed with an exact value. It can only be known that it is around 1 GeV in QCD from the phenomenological point of view. It is indeed needed to study the behavior of the physical results varying with the value of the critical scale $\mu_c$. We study this effect by slightly varying the value of $\mu_c$ around 1 GeV with fixing the physical $B\pi$, $BK$ transition form factors, the total $K\pi$ production from factor and the scale-independence of the contribution of the color-octet hadronic matrix element. The result is shown in Table \ref{varyingscale}. It is shown in the table that the decay branching ratios and $CP$ violations are not changed very much when varying $\mu_c$ from 0.9$\sim$ 1.3 GeV. The change becomes apparent only when $\mu_c>2.0\;\mathrm{GeV}$, where the soft contribution has been pushed to a scale of too high. Therefore, $\mu_c=1.0\;\mathrm{GeV}$ is an acceptable choice for phenomenological study. Certainly, it can be varied slightly around 1 GeV.

3) The introduction of soft quantities, such as the soft form factors and color-octet contribution, changes the power counting rule for the decay amplitudes in $B$ decays. It is different from both PQCD approach in the early stage \cite{PQCD1,PQCD2,PQCD3} and QCDF approach \cite{beneke1999qcd,beneke2000qcd,beneke2001qcd}. These soft quantities fully contribute at the soft scale $\mu_c$, which are crucial to diminish the tension between the theoretical predictions and experimental data on $B\to K\pi$ decays, and the physical results showed a bit of stability when varying the critical cutoff scale $\mu_c$ around 1 GeV.

\begin{table*}
  \caption{\label{varyingscale}
    $B\rightarrow K\pi$ branching ratios and \textit{CP} violations varying with the critical cutoff scale $\mu_c$, where the total from factors are fixed with $F_0^{BK}(0)=0.33$, $F_0^{B\pi}(0)=0.27$ and$F_+^{K\pi}=0.20\exp(-0.47i\pi)$
    \footnote{The total form factors should not vary with the critical cutoff scale. So the value of them can be obtained by adding the hard and soft part at any value of $\mu_c$. Here $F_+^{K\pi}$ is taken by adding the values of $h^{K\pi}$ and $\xi^{K\pi}$ at $\mu_c=1\;\mathrm{GeV}$. The color-octet contributions are taken as $\mu_c $-independent quantities.}}
  \renewcommand\arraystretch{1.5}
  \begin{ruledtabular}
    \begin{tabular}{ccccccccc}
      Mode
      &$\mu_c=0.9\mathrm{GeV}$&$1.0\mathrm{GeV}$ &$1.1\mathrm{GeV}$ &$1.3\mathrm{GeV}$  &$1.5\mathrm{GeV}$ &$2.0\mathrm{GeV}$
      & {Data} \cite{Zyla:2020zbs} \\
      \hline
      $B({B}^+\rightarrow K^0\pi^+)\times 10^{-6}$& $24.5$ & $24.3$ & $24.3$ & $23.5$ & $23.0$ & $20.9$ &$23.7\pm 0.8$ \\
      $B({B}^+\rightarrow K^+\pi^0)\times 10^{-6}$& $12.4$ & $12.6$ & $12.7$ & $12.5$ & $12.4$ & $11.5$ &$12.9\pm 0.5$ \\
      $B({B}^0\rightarrow K^+\pi^-)\times 10^{-6}$& $19.5$ & $20.0$ & $20.4$ & $20.5$ & $20.4$ & $19.0$ &$19.6\pm 0.5$ \\
      $B({B}^0\rightarrow K^0\pi^0)\times 10^{-6}$& $ 9.3$ & $ 9.4$ & $ 9.5$ & $ 9.3$ & $ 9.1$ & $ 8.4$ &$9.9 \pm 0.5$ \\
      $A_{CP}(B^+\rightarrow K^0\pi^+)$           &$ 0.012$&$ 0.011$&$ 0.011$&$ 0.010$&$ 0.009$&$ 0.008$&$-0.017\pm 0.016$ \\
      $A_{CP}(B^+\rightarrow K^+\pi^0)$           &$ 0.055$&$ 0.041$&$ 0.031$&$ 0.012$&$ 0.001$&$-0.011$&$ 0.037\pm 0.021$ \\
      $A_{CP}(B^0\rightarrow K^+\pi^-)$           &$-0.055$&$-0.084$&$-0.102$&$-0.133$&$-0.150$&$-0.178$&$-0.083\pm 0.004$ \\
      $A_{CP}(B^0\rightarrow K^0\pi^0)$           &$-0.100$&$-0.112$&$-0.118$&$-0.128$&$-0.133$&$-0.148$&$ 0.00 \pm 0.13$ \\
    \end{tabular}
  \end{ruledtabular}
\end{table*}

\section{summary} \label{sec:sum}
In this work we study the $B\rightarrow K\pi$ decays in a modified perturbative QCD approach. With the $B$ meson wave function that is obtained in the relativistic potential model, we find that soft contribution can not be suppressed enough by Sudakov factor. It is necessary to introduce the soft scale cutoff and soft form factors.
In addition, we also introduce the hadronic color-octet matrix element which plays an important role in explaining the dramatic difference between the \textit{CP} violations of $B^+\rightarrow K^+\pi^0$ and $B^0\rightarrow K^+\pi^-$ decays. Taking appropriate values for the input parameters, our calculated results for all the branching ratios and most \textit{CP} violations of the $B\rightarrow K\pi$ decay channels are well consistent with the experimental data, except for the \textit{CP} violation in the $B^+\rightarrow K^0\pi^+$ decay mode, which is very close to the experimental data.

\begin{acknowledgments}
  This work is supported in part by the National Natural Science Foundation of China
  under Contracts No. 11875168, 12275139.
\end{acknowledgments}

\vspace{0.5cm}

\appendix

\section{Formulas in the Hard Part Calculations} \label{ap:factors}
The threshold factor $S_t(x)$ is usually parameterized as \cite{li2002threshold}
\begin{equation}
  S_t(x)=\frac{2^{1+2c}\Gamma(3/2+c)}{\sqrt{\pi}\Gamma(1+c)}[x(1-x)]^c,
\end{equation}
with $c=0.3$.

The exponentials $\exp[-S_{B,K,\pi}(t)]$ include the Sudakov factor and single ultraviolet logarithms
which is related to the meson wave functions.
The expressions of exponents are
\begin{equation}
  \begin{split}
    S_B(t)=&s(x_1,b_1,m_B)-\frac{1}{\beta_1}
      \ln\frac{\ln(t/\Lambda_{\textup{QCD}})}{\ln(1/(b_1\Lambda_{\textup{QCD}}))}, \\
  \end{split}
\end{equation}
\begin{equation}
  \begin{split}
    S_K(t)=&s(x_2,b_2,m_B)+s(1-x_2,b_2,m_B) \\
    &-\frac{1}{\beta_1}\ln\frac{\ln(t/\Lambda_{\textup{QCD}})}{\ln(1/(b_2\Lambda_{\textup{QCD}}))}, \\
  \end{split}
\end{equation}
\begin{equation}
  \begin{split}
    S_\pi(t)=&s(x_3,b_3,m_B)+s(1-x_3,b_3,m_B) \\
    &-\frac{1}{\beta_1}\ln\frac{\ln(t/\Lambda_{\textup{QCD}})}{\ln(1/(b_3\Lambda_{\textup{QCD}}))}. \\
  \end{split}
\end{equation}
The explicit form of $s(x,b,Q)$ up to next-to-leading order is \cite{li1995applicability}
\begin{widetext}
\begin{equation} \label{eq:sudakov}
  \begin{split}
    s(x,b,Q)=&\frac{A^{(1)}}{2\beta_1}\hat{q}\ln\left(\frac{\hat{q}}{\hat{b}}\right)
      -\frac{A^{(1)}}{2\beta_1}\left(\hat{q}-\hat{b}\right)
      +\frac{A^{(2)}}{4\beta_1^2}\left(\frac{\hat{q}}{\hat{b}}-1\right)
      -\left[\frac{A^{(2)}}{4\beta_1^2}-\frac{A^{(1)}}{4\beta_1}
      \ln\left(\frac{e^{2\gamma_E-1}}{2}\right)\right]
      \ln\left(\frac{\hat{q}}{\hat{b}}\right) \\
    &+\frac{A^{(1)}\beta_2}{4\beta_1^3}\hat{q}
      \left[\frac{\ln(2\hat{q})+1}{\hat{q}}-\frac{\ln(2\hat{b})+1}{\hat{b}}\right]
      +\frac{A^{(1)}\beta_2}{8\beta_1^3}
      \left[\ln^2(2\hat{q})-\ln^2(2\hat{b})\right] \\
    &+\frac{A^{(1)}\beta_2}{8\beta_1^3}
      \ln\left(\frac{e^{2\gamma_E-1}}{2}\right)
      \left[\frac{\ln(2\hat{q})+1}{\hat{q}}-\frac{\ln(2\hat{b})+1}{\hat{b}}\right] \\
    &-\frac{A^{(1)}\beta_2}{16\beta_1^4}
      \left[\frac{2\ln(2\hat{q})+3}{\hat{q}}-\frac{2\ln(2\hat{b})+3}{\hat{b}}\right]
      -\frac{A^{(1)}\beta_2}{16\beta_1^4}
      \frac{\hat{q}-\hat{b}}{\hat{b}^2}\left[2\ln(2\hat{b})+1\right] \\
    &+\frac{A^{(2)}\beta_2^2}{1728\beta_1^6}
      \left[\frac{18\ln^2(2\hat{q})+30\ln(2\hat{q})+19}{\hat{q}^2}
      -\frac{18\ln^2(2\hat{b})+30\ln(2\hat{b})+19}{\hat{b}^2}\right] \\
    &+\frac{A^{(2)}\beta_2^2}{432\beta_1^6}\frac{\hat{q}-\hat{b}}{\hat{b}^3}
      \left[9\ln^2(2\hat{b})+6\ln(2\hat{b})+2\right], \\
  \end{split}
\end{equation}
\end{widetext}
where $\hat{q}$ and $\hat{b}$ are defined as
\begin{equation}
  \hat{q}\equiv \ln\left(xQ/(\sqrt{2}\Lambda_{\textup{QCD}})\right),\quad
  \hat{b}\equiv \ln\left(1/{b\Lambda_{\textup{QCD}}}\right).
\end{equation}
The coefficients $\beta_i$ and $A^{(i)}$ in Eq.~\eqref{eq:sudakov} are
\begin{equation}
    \beta_1=\frac{33-2n_f}{12}, \quad
    \beta_2=\frac{153-19n_f}{24},
\end{equation}
and
\begin{equation}
  \begin{split}
    A^{(1)}=&\frac{4}{3}, \\
    A^{(2)}=&\frac{67}{9}-\frac{\pi^2}{3}
      -\frac{10}{27}n_f+\frac{8}{3}\beta_1\ln\left(\frac{e^{\gamma_E}}{2}\right), \\
  \end{split}
\end{equation}
where $\gamma_E$ is Euler constant.

In the Eqs.~\eqref{eq:fe}--\eqref{eq:map}, the function $h$'s are given as
\begin{widetext}
\begin{equation}
  \begin{split}
    h_e(x_1,x_2,b_1,b_2)=&K_0(\sqrt{x_1x_2}m_Bb_1)
      \Bigl[\theta(b_1-b_2)K_0(\sqrt{x_2}m_Bb_1)I_0(\sqrt{x_2}m_Bb_2) \\
    &+\theta(b_2-b_1)K_0(\sqrt{x_2}m_Bb_2)I_0(\sqrt{x_2}m_Bb_1)\Bigr], \\
  \end{split}
\end{equation}
\begin{equation}
  \begin{split}
    h_a(x_1,x_2,b_1,b_2)=&K_0(-i\sqrt{x_1x_2}m_Bb_1)
      \Bigl[\theta(b_1-b_2)K_0(-i\sqrt{x_2}m_Bb_1)I_0(-i\sqrt{x_2}m_Bb_2) \\
    &+\theta(b_2-b_1)K_0(-i\sqrt{x_2}m_Bb_2)I_0(-i\sqrt{x_2}m_Bb_1)\Bigr], \\
  \end{split}
\end{equation}
\begin{equation}
  \begin{split}
    h_d(x_1,x_2,x_3,b_1,b_2)=&K_0(-i\sqrt{x_2x_3}m_Bb_2)
      \Bigl[\theta(b_1-b_2)K_0(\sqrt{x_1x_3}m_Bb_1)I_0(\sqrt{x_1x_3}m_Bb_2)\\
    &+\theta(b_2-b_1)K_0(\sqrt{x_1x_3}m_Bb_2)I_0(\sqrt{x_1x_3}m_Bb_1)\Bigr],\\
  \end{split}
\end{equation}
\begin{equation}
  \begin{split}
    h_f^1(x_1,x_2,b_1,b_2)=&K_0(-i\sqrt{x_1x_2}m_Bb_1)
      \Bigl[\theta(b_1-b_2)K_0(-i\sqrt{x_1x_2}m_Bb_1)I_0(-i\sqrt{x_1x_2}m_Bb_2)\\
    &+\theta(b_2-b_1)K_0(-i\sqrt{x_1x_2}m_Bb_2)I_0(-i\sqrt{x_1x_2}m_Bb_1)\Bigr],\\
  \end{split}
\end{equation}
\begin{equation}
  \begin{split}
    h_f^2(x_1,x_2,b_1,b_2)=&K_0(\sqrt{x_1+x_2-x_1x_2}m_Bb_1)
      \Bigl[\theta(b_1-b_2)K_0(-i\sqrt{x_1x_2}m_Bb_1)I_0(-i\sqrt{x_1x_2}m_Bb_2)\\
    &+\theta(b_2-b_1)K_0(-i\sqrt{x_1x_2}m_Bb_2)I_0(-i\sqrt{x_1x_2}m_Bb_1)\Bigr],\\
  \end{split}
\end{equation}
where $J_0$, $K_0$ and $I_0$ are Bessel and modified Bessel functions.
\end{widetext}

\section{Light Meson Distribution Amplitudes} \label{ap:kaonwf}
The transverse momentum dependence of $\phi_M(x,k_\perp)$, $\phi_P^M(x,k_\perp)$
and $\phi_\sigma^M(x,k_\perp)$ is assumed to be a Gaussian distribution,
where $M=\pi,K$.
When transform the wave function into $b$-space, the distribution amplitudes become
\begin{equation}
    \phi(x,b)=\phi(x)\exp(-\frac{b^2}{4\beta^2}),
\end{equation}
for $\phi_M(x,b)$, $\phi_P^M(x,b)$ and $\phi_\sigma^M(x,b)$.
The oscillation parameter $\beta$ can be related to the root mean square transverse momentum by
$\beta=1/\sqrt{2\langle k_\perp^2\rangle}$ \cite{wy2002}.
The reasonable value of the root mean square transverse momentum for pion is 350 MeV
according to the study of pion form factor in Ref. \cite{JK93},
which is relevant to $\beta=4.0 ~\mathrm{GeV}^{-1} $.
Here we take $\beta=4.0 \;\mathrm{GeV}^{-1} $ for both pion and kaon's wave functions.
The twist-2 and twist-3 distribution amplitudes $\phi_M(x)$, $\phi_P^M(x)$ and
$\phi_\sigma^M(x)$ are given by \cite{ball2006higher}
\begin{equation}
  \label{eq:phi}
  \phi_M(x)=6x(1-x)\biggl[1+a_1^M C_1^{3/2}(t)+a_2^M C_2^{3/2}(t)\biggr],
\end{equation}
\begin{equation}
  \label{eq:phip}
  \begin{split}
    \phi_P^M(x)&=1+a_{0P}^M+a_{1P}^MC_1^{1/2}(t)+a_{2P}^MC_2^{1/2}(t) \\
      &\quad+a_{3P}^MC_3^{1/2}(t) +a_{4P}^MC_4^{1/2}(t) \\
      &\quad+b_{1P}^M\ln(x)+b_{2P}^M\ln(1-x),\\
  \end{split}
\end{equation}
\begin{equation}
  \label{eq:phis}
  \begin{split}
    \phi_\sigma^M(x)&=6x(1-x)\biggl[1+a_{0\sigma}^M+a_{1\sigma}^MC_1^{3/2}(t) \\
      &\quad+a_{2\sigma}^MC_2^{3/2}(t)+a_{3\sigma}^MC_3^{3/2}(t)\biggr] \\
      &\quad+9x(1-x)\biggl[b_{1\sigma}^M\ln(x)+b_{2\sigma}^M\ln(1-x)\biggr],\\
  \end{split}
\end{equation}
where $t=2x-1$.
The function $C$'s are Gegenbauer polynomials.
The coefficients $a_{i(P,\sigma)}^M$ and $b_{i(P,\sigma)}^M$ in Eqs.~\eqref{eq:phi}--\eqref{eq:phis} are
\begin{equation}
  \begin{split}
    &a_1^\pi=0, \quad a_2^\pi=0.25\pm 0.15, \\
    &a_{0P}^\pi=0.048\pm 0.017, \quad a_{2P}^\pi=0.62\pm 0.20,\\
    &a_{4P}^\pi=0.089\pm 0.051, \quad a_{1P}^\pi=a_{3P}^\pi=0, \\
    &b_{1P}^\pi=b_{2P}^\pi=0.024\pm 0.009, \\
    &a_{0\sigma}^\pi=0.034\pm 0.014,\quad a_{2\sigma}^\pi=0.12\pm 0.03, \\
    &a_{1\sigma}^\pi=a_{3\sigma}^\pi=0, \quad b_{1\sigma}^\pi=b_{2\sigma}^\pi=0.016\pm 0.006, \\
\end{split}
\end{equation}
for pion, and
\begin{equation}
  \begin{split}
    &a_1^K=0.06\pm 0.03, \quad a_2^K=0.25\pm 0.15, \\
    &a_{0P}^K=0.58\pm 0.23, \quad a_{1P}^K=-0.57\pm 0.31, \\
&a_{2P}^K=0.79\pm 0.25, \quad a_{3P}^K=0.18\pm 0.12, \\
    &a_{4P}^K=0.06\pm 0.04, \\
\end{split}
\end{equation}
\begin{equation}
  \begin{split}
    &b_{1P}^K=0.56\pm 0.22, \quad b_{2P}^K=0.03\pm 0.01, \\
    &a_{0\sigma}^K=0.40\pm 0.19, \quad a_{1\sigma}^K=-0.13\pm 0.09, \\
    &a_{2\sigma}^K=0.12\pm 0.03, \quad a_{3\sigma}^K=0.03\pm 0.01, \\
    &b_{1\sigma}^K=0.37\pm 0.14, \quad b_{2\sigma}^K=0.02\pm 0.01, \\
  \end{split}
\end{equation}
for kaon. All the above parameters are determined at the renormalization scale $\mu=1.0~\mathrm{GeV}$.
 The Gegenbauer polynomials are given by
 \begin{equation}
   \begin{split}
     &C_1^{1/2}(t)=t,\\
     &C_2^{1/2}(t)=\frac{1}{2}\left(3t^2-1\right), \\
     &C_3^{1/2}(t)=\frac{t}{2}\left(5t^2-3\right), \\
     &C_4^{1/2}(t)=\frac{1}{8}\left(35t^4-30t^2+3\right), \\
   \end{split}
 \end{equation}
 \begin{equation}
   \begin{split}
     &C_1^{3/2}(t)=3t, \\
     &C_2^{3/2}(t)=\frac{3}{2}\left(5t^2-1\right), \\
     &C_3^{3/2}(t)=\frac{5}{2}t\left(7t^2-3\right), \\
     &C_4^{3/2}(t)=\frac{15}{8}\left(21t^4-14t^2+1\right). \\
   \end{split}
 \end{equation}

\end{document}